\documentclass[aps,prb,twocolumn,groupedaddress,superscriptaddress,nobibnotes]{revtex4-2}
\usepackage{amsmath}
\usepackage{amssymb}
\usepackage{graphicx}
\usepackage{dcolumn}
\usepackage{bm}
\usepackage{xcolor}
\usepackage[normalem]{ulem}
\usepackage{siunitx}
\usepackage{soul}
\setulcolor{red}

\begin{document}

\title{Insight into the electronic structure of the centrosymmetric skyrmion magnet {GdRu$_2$Si$_2$}}

\author{S.\,V. Eremeev}
\email{eremeev@ispms.tsc.ru} \affiliation{Institute of Strength
Physics and Materials Science, Russian Academy of Sciences, 634055
Tomsk, Russia}

\author{D.~Glazkova}
\affiliation{St. Petersburg State University, 7/9 Universitetskaya
nab., St. Petersburg, 199034, Russia}

\author{G.~Poelchen}
\affiliation{Institut f\"ur Festk\"orper- und Materialphysik,
Technische Universit\"{a}t Dresden, D-01062 Dresden, Germany}

\author{A.~Kraiker}
\affiliation{Kristall- und Materiallabor, Physikalisches Institut,
Goethe-Universit\"{a}t Frankfurt, Max-von-Laue Strasse 1, D-60438
Frankfurt am Main, Germany}

\author{K.~Ali}
\affiliation{Department of Microtechnology and Nanoscience, Chalmers
University of Technology, G\"oteborg, 41296 Sweden}

\author{A.~V.~Tarasov}
\affiliation{St. Petersburg State University, 7/9 Universitetskaya
nab., St. Petersburg, 199034, Russia}

\author{S.~Schulz}
\affiliation{Institut f\"ur Festk\"orper- und Materialphysik,
Technische Universit\"{a}t Dresden, D-01062 Dresden, Germany}

\author{K.~Kliemt}
\affiliation{Kristall- und Materiallabor, Physikalisches Institut,
Goethe-Universit\"{a}t Frankfurt, Max-von-Laue Strasse 1, D-60438
Frankfurt am Main, Germany}

\author{E.V.~Chulkov}
 \affiliation{Departamento de Pol\'imeros y Materiales Avanzados: F\'isica, Qu\'imica y Tecnolog\'ia, Facultad de Ciencias Qu\'imicas, Universidad del Pa\'is Vasco UPV/EHU, 20080 San Sebasti\'an/Donostia, Spain}
 \affiliation{Centro de F\'isica de Materiales (CFM-MPC), Centro Mixto CSIC-UPV/EHU, 20018 San Sebasti\'{a}n/Donostia, Spain}
  \affiliation{Donostia International Physics Center (DIPC), 20018 Donostia-San Sebasti\'{a}n, Spain}
 \affiliation{St. Petersburg State University, 7/9 Universitetskaya nab., St. Petersburg, 199034, Russia}

\author{V.~S.~Stolyarov}
\affiliation{Moscow Institute of Physics and Technology, Institute
Lane 9, Dolgoprudny, Russia} \affiliation{Dukhov Research Institute
of Automatics (VNIIA), Moscow, 127055 Russia} \affiliation{National
University of Science and Technology MISIS, Moscow, 119049 Russia}

\author{A.~Ernst}
\affiliation{Institute for Theoretical Physics, Johannes Kepler
University, Linz, Austria}

\author{C.~Krellner}
\affiliation{Kristall- und Materiallabor, Physikalisches Institut,
Goethe-Universit\"{a}t Frankfurt, Max-von-Laue Strasse 1, D-60438
Frankfurt am Main, Germany}

\author{D.Yu. Usachov}
\affiliation{St. Petersburg State University, 7/9 Universitetskaya
nab., St. Petersburg, 199034, Russia} \affiliation{Moscow Institute
of Physics and Technology, Institute Lane 9, Dolgoprudny, Russia}
\affiliation{National University of Science and Technology MISIS,
Moscow, 119049 Russia}

\author{D.V. Vyalikh}
\email{denis.vyalikh@dipc.org} \affiliation{Donostia International
Physics Center (DIPC), 20018 Donostia-San Sebasti\'{a}n, Spain}
\affiliation{IKERBASQUE, Basque Foundation for Science, 48011
Bilbao, Spain}

\begin{abstract}
The discovery of a square magnetic-skyrmion lattice in
GdRu$_2$Si$_2$, with the smallest so far found skyrmion diameter and
without a geometrically frustrated lattice, has attracted
significant attention, particularly for potential applications in
memory devices and quantum computing. In this work, we present a
comprehensive study of surface and bulk electronic structures of
GdRu$_2$Si$_2$ by utilizing momentum-resolved photoemission (ARPES)
measurements and first-principles calculations. We show how the
electronic structure evolves during the antiferromagnetic transition
when a peculiar helical order of 4$f$ magnetic moments within the Gd
layers sets in. A nice agreement of the ARPES-derived electronic
structure with the calculated one has allowed us to characterize the
features of the Fermi surface (FS), unveil the nested region along
the $k_z$ at the corner of the 3D FS, and reveal their orbital
compositions. Our findings suggest that the
Ruderman-Kittel-Kasuya-Yosida interaction plays a decisive role in
stabilizing the spiral-like order of Gd 4$f$ moments responsible for
the skyrmion physics in GdRu$_2$Si$_2$. Our results provide a deeper
understanding of electronic and magnetic properties of this
material, which is crucial for predicting and developing novel
skyrmion-based devices.
\end{abstract}

\maketitle

\section{Introduction}

Known since the beginning of the 80s, the centrosymmetric
antiferromagnet GdRu$_2$Si$_2$ ($T_N$ of $\sim$ 46 K) which
crystallizes in the ThCr$_2$Si$_2$ structure with I4/mmm space
symmetry~\cite{Slaski_JLCM1982,Hiebl_JMMM1983,Slaski_JMMM1984} has
recently reappeared in the focus of research efforts due to the
discovery of a square magnetic-skyrmion lattice without a
geometrically frustrated lattice~\cite{Khanh_NatNano2020}. This
skyrmion phase appears in an external magnetic field of 2--2.5 T at
temperatures below 20 K. Although magnetic properties of
GdRu$_2$Si$_2$ have been investigated over the years very
extensive~\cite{Slaski_JLCM1982,Hiebl_JMMM1983,Slaski_JMMM1984,Garnier_JMMM1995,Rotter_JMMM2007,Samanta_JAP2008,Garnier_PB1996,Prokleska_JP2006}
the emergence of the skyrmion phase has renewed and intensified the
discussions about this material, specifically regarding the reason
for and the origin of the skyrmion
physics~\cite{Khanh_NatNano2020,Bouaziz_PRL2022,Yasui_NatComm2020,Nomoto_PRL2020,Hayami_PRB2021}.
It is worth noting that the discovered square skyrmion lattice has,
with a diameter of \SI{1.9}{nm}, the smallest skyrmion size ever
observed, which makes it attractive for development of
next-generation high-density magnetic memory devices with ultralow
energy consumption~\cite{Khanh_NatNano2020,Yasui_NatComm2020}.

For further studies of the properties of this material and
prediction of possible candidates which could reveal unusual
magnetic-skyrmion properties as well, detailed information about
surface and bulk electronic structures and, most importantly, on how
the electronic structure gets modified upon the antiferromagnetic
transition is highly desirable. The latter can be derived from
momentum-resolved photoemission (ARPES) measurements combined with
\emph{ab initio} calculations. The curious question here is in how
far it is possible to detect the stabilization of the helical
long-range in-plane magnetic order of Gd 4$f$ moments appearing at
zero magnetic field in the ARPES patterns. The next essential
question is about the Fermi surface (FS) and its intrinsic
properties. Recently, contradicting results were presented for the
properties of the three-dimensional (3D) FS. In a recent theoretical
study \cite{Bouaziz_PRL2022}, the nested region near the
$\Gamma$-point was manifested and it was declared to be responsible
for the skyrmion physics in GdRu$_2$Si$_2$. However, results of
Ref.~\citenum{Matsuyama_PRB23}, which also present a detailed
overview of the actual discussion about the properties of
GdRu$_2$Si$_2$, do not confirm the nested region near the
$\Gamma$-point. This makes a call for a direct experimental
visualization of the Fermi surface and further exploration of its
properties via ARPES measurements.

Here, we present a detailed study of the electronic structure and
discuss FS properties for GdRu$_2$Si$_2$ obtained from ARPES
measurements and \emph{ab initio} density functional theory (DFT)
calculations. The latter have been performed for the bulk and for a
finite system (slab) that allows us to distinguish surface
resonances and surface states for different surface terminations and
separate them from the bulk states. Note that most of the surface
electron states were observed for the Si-terminated surface of
GdRu$_2$Si$_2$, where the topmost Gd layer is hidden by the Si-Ru-Si
surface trilayer block. In our study, this surface will be of the
main focus.

First, our analysis of GdRu$_2$Si$_2$ will be focused on the
characterization of the electronic structure in the paramagnetic
phase, which shows a good agreement between experiment and theory.
In the AFM ordered phase, our ARPES measurements reveal a pseudogap
which is seen as a kind of a "sickle-moon" feature near the Fermi
level ($E_\mathrm{F}$) in the structure of the bulk projected bands
taken along the
$\overline{\mathrm{M}}-\overline{\mathrm{X}}-\overline{\mathrm{M}}$
direction of the surface Brillouin zone (BZ). Our theoretical
analysis indicates that this feature is directly linked with the
formation of a spiral structure of Gd 4$f$ moments within the basal
plane. Further, the agreement between the ARPES-derived electronic
structure and the one based on DFT calculations enables a
comprehensive characterization of bulk and surface states, including
their properties and orbital compositions. Here, we found that in
contrast to earlier predictions the FS near the $\Gamma$-point does
not exhibit nesting properties. Instead, a nesting occurs near the
corner of the BZ with a vector along [100]([010]), which connects
$k_z$-dispersionless states with noticeable Gd-$d$ contribution. The
size of the nesting vector fits well to describe the properties of
the spiral magnetic structure. This indicates that the electronic
states in this BZ region are most essential for a RKKY-mediated
coupling of Gd 4$f$ moments. These results are inline with those
recently reported in Ref.~\citenum{Matsuyama_PRB23}. Lastly, we
discuss results of a theoretical modelling with stretching of the
unit cell within the basal plane, analyzing the modifications of the
corner-nested region of the BZ. The obtained results allow us to
propose the most plausible scenario for the emergence of the helical
AFM phase in GdRu$_2$Si$_2$.

\section{Results}

\subsection{Calculated electronic structure of the paramagnetic GdRu$_2$Si$_2$ (001) surfaces}

\begin{figure*}
  \includegraphics[width=\textwidth]{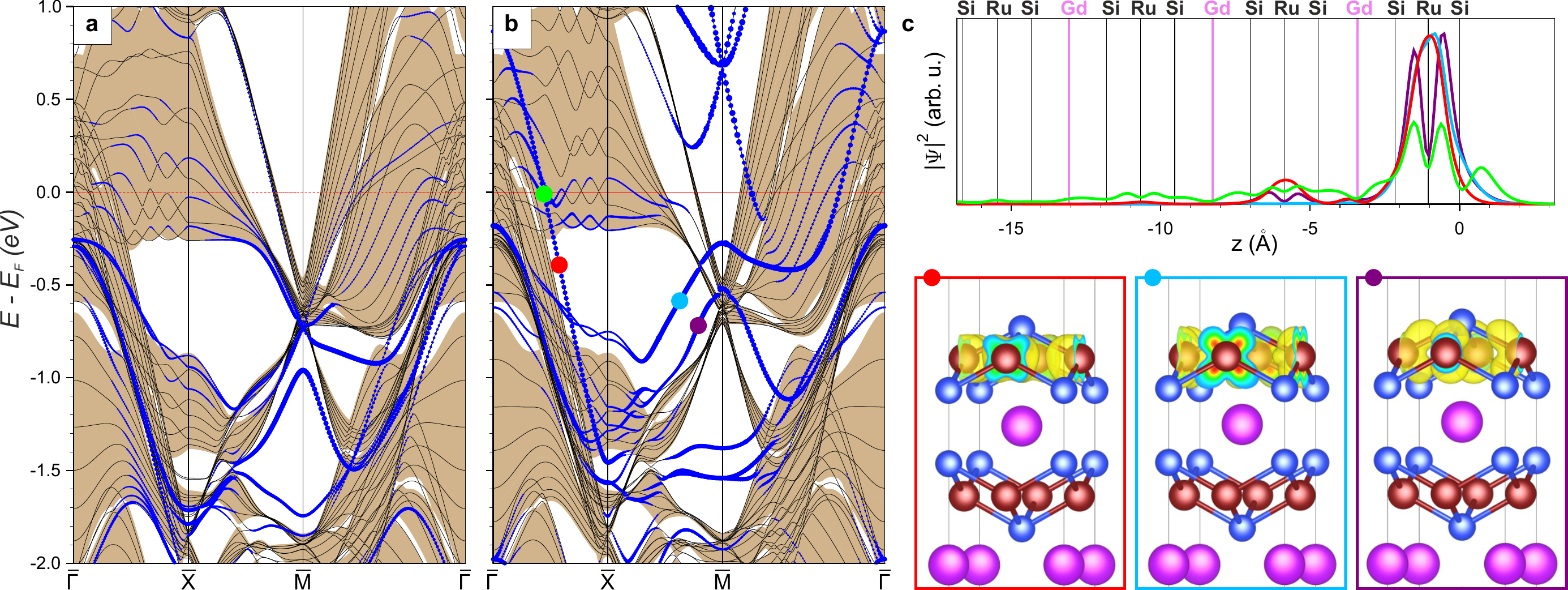}
  \caption{Electronic structure of the paramagnetic phase of GdRu$_2$Si$_2$(001) calculated for Gd- (a) and Si-terminated (b) surfaces. The tan-shaded area corresponds to the surface-projected bulk band structure, whereas the black lines are the result of a slab model. The size of blue dots reflect the weights of the surface states. Large colored circles in (b) mark the states whose spatial distribution is shown in (c) in the same colors. (c) Upper panel: Electron density distribution (integrated over the $ab$ plane) of the surface electron states marked in (b). Vertical lines reveal the positions of Gd (pink) and Si-Ru-Si (black) atomic layers. Lower panel: spatial distributions for the ``red", ``light blue", and ``purple" surface states.
  }
  \label{fig1}
\end{figure*}

We begin with a theoretical analysis of the bulk and surface
electronic structure of the paramagnetic GdRu$_2$Si$_2$ obtained
from \emph{ab initio} DFT calculations. Similarly to many others
RET$_2$Si$_2$ materials~\cite{Gustav2022}, the predominant cleavage
plane of single crystal of GdRu$_2$Si$_2$ (001) is between Si- and
Gd- atomic layers~\cite{Yasui_NatComm2020}, leaving behind an either
Si- or Gd-terminated surface. Therefore, to determine surface
electron states and separate them from bulk electron bands, an
asymmetric slab terminated by Si and Gd on either side was used
allowing us to trace bulk-like bands, band gaps as well as
surface-related states for both terminations. The in this way
theoretically derived electron states for the Gd and Si surfaces of
the GdRu$_2$Si$_2$(001) are shown in Fig.~\ref{fig1}(a, b),
respectively.  The surface electron states are shown in blue
overlaid with the tan-colored $k_z$-projected bulk states. The
strongly localized Gd $4f$ orbitals were treated as a frozen core
approximation.

Both surfaces reveal a number of surface-related electron states.
While some of them overlap strongly with the bulk bands, others
reside within large projected band gaps. As we will see further, the
theoretically-derived spectral patterns agree nicely with those
obtained in the momentum-resolved photoemission experiment. The
in-gap surface states seen for the Si surface are almost fully
localized within the topmost trilayer Si-Ru-Si block and are mainly
composed of Ru $4d$ orbitals with different symmetries (lower panel
of Fig.~\ref{fig1}c). We mark a few examples of such states by the
colored (red, lightblue and purple) circles in Fig.~\ref{fig1}b and
present their density profiles $\left|\Psi\right|^2$ in the upper
panel of Fig.~\ref{fig1}c. The lack of their overlap with the Gd
layer makes these surface states rather insensitive to the
stabilization of a spin-spiral magnetic order in the AFM phase of
GdRu$_2$Si$_2$. However, other surface states considerably overlap
with the bulk projected states and penetrate into their continuum.
Hence, it is reasonable to anticipate that those states, like the
one labelled by in green situated between $\overline{\Gamma}$ and
$\overline{\mathrm{X}}$ close to the $E_\mathrm{F}$, should get
modified when the exotic magnetic order on the Gd sublattice sets
in. We will discuss the sensitivity of these states to the magnetic
order in detail later in comparison with our experimental ARPES
results.

\subsection{ARPES}

\begin{figure*}
  \includegraphics[width=\textwidth]{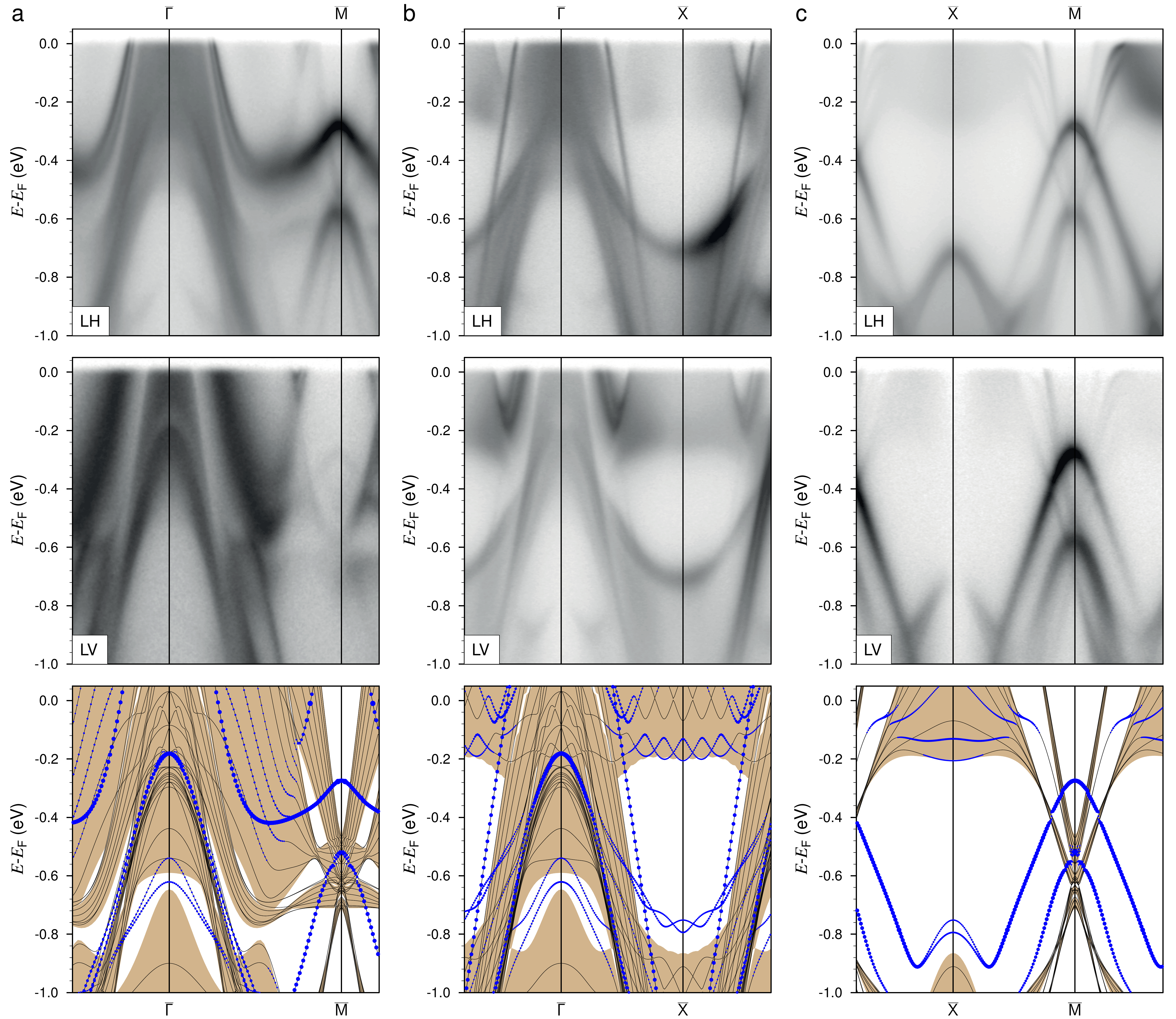}
  \caption{Comparison between ARPES spectra and DFT-derived band structures for the paramagnetic Si-terminated (001) surface of GdRu$_2$Si$_2$ along $\overline{\Gamma}-\overline{\mathrm{M}}$ (a), $\overline\Gamma-\overline{\mathrm{X}}$ (b), and $\overline{\mathrm{X}}-\overline{\mathrm{M}}$ (c) directions measured with linear horizontal (LH) (top row) and linear vertical (LV) (middle row) light polarizations. Bottom row presents the DFT-derived band structure shown in the same energy window for better comparison with ARPES data.
  }
  \label{fig2}
\end{figure*}

To comprehensively explore the complex electronic structure of the
paramagnetic GdRu$_2$Si$_2$, we performed detailed ARPES
measurements and compared the obtained results with those derived
from our first principle calculations. In the first two rows of
Fig.~\ref{fig2}, we present the experimentally derived electronic
structure taken from the Si-termination of a freshly cleaved
GdRu$_2$Si$_2$ single crystal at \SI{50}{K} along the
$\overline{\Gamma}-\overline{\mathrm{M}}$,
$\overline\Gamma-\overline{\mathrm{X}}$ and
$\overline{\mathrm{X}}-\overline{\mathrm{M}}$ directions of the
surface BZ. The ARPES pattern were taken at a photon energy of
\SI{50}{eV} with horizontal (LH) and vertical (LV) polarizations
shown in the first and second row of Fig.~\ref{fig2}, respectively.
In the third row, we display the $k_z$-projected bulk states along
the same high-symmetry directions as in experiment. Similarly to
Fig.~\ref{fig1}, the tan-colored bulk states are overlaid with the
slab-derived surface and surface-resonant states depicted in blue
for the paramagnetic, Si-terminated  GdRu$_2$Si$_2$ (001) surface.

The good agreement between experiment and theory allows us to easily
identify and differentiate between the surface-related bands and the
bulk-projected electronic structure in the ARPES patterns. This
gives us a good starting point for the subsequent evaluation of how
those states evolve when the peculiar spiral magnetic order of the
Gd 4$f$ moments stabilizes in the AFM phase of GdRu$_2$Si$_2$.
First, we will focus on the bulk states close to $E_\mathrm{F}$,
their properties and orbital composition. Based on
Ref.~\citenum{Bouaziz_PRL2022}, special attention will be paid to
the areas near high-symmetry points and in particular on the center
of the BZ. Secondly, we will investigate the behavior of the
surface-related states which overlap with the bulk states (labelled
in green in Fig.~\ref{fig1}b). Lastly, we can then discuss and
analyse the whole Fermi surface as derived from ARPES measurements
and DFT calculations.

\subsection{Effect of the spin spiral on the electronic structure of GdRu$_2$Si$_2$}

\begin{figure*}
  \includegraphics[width=\textwidth]{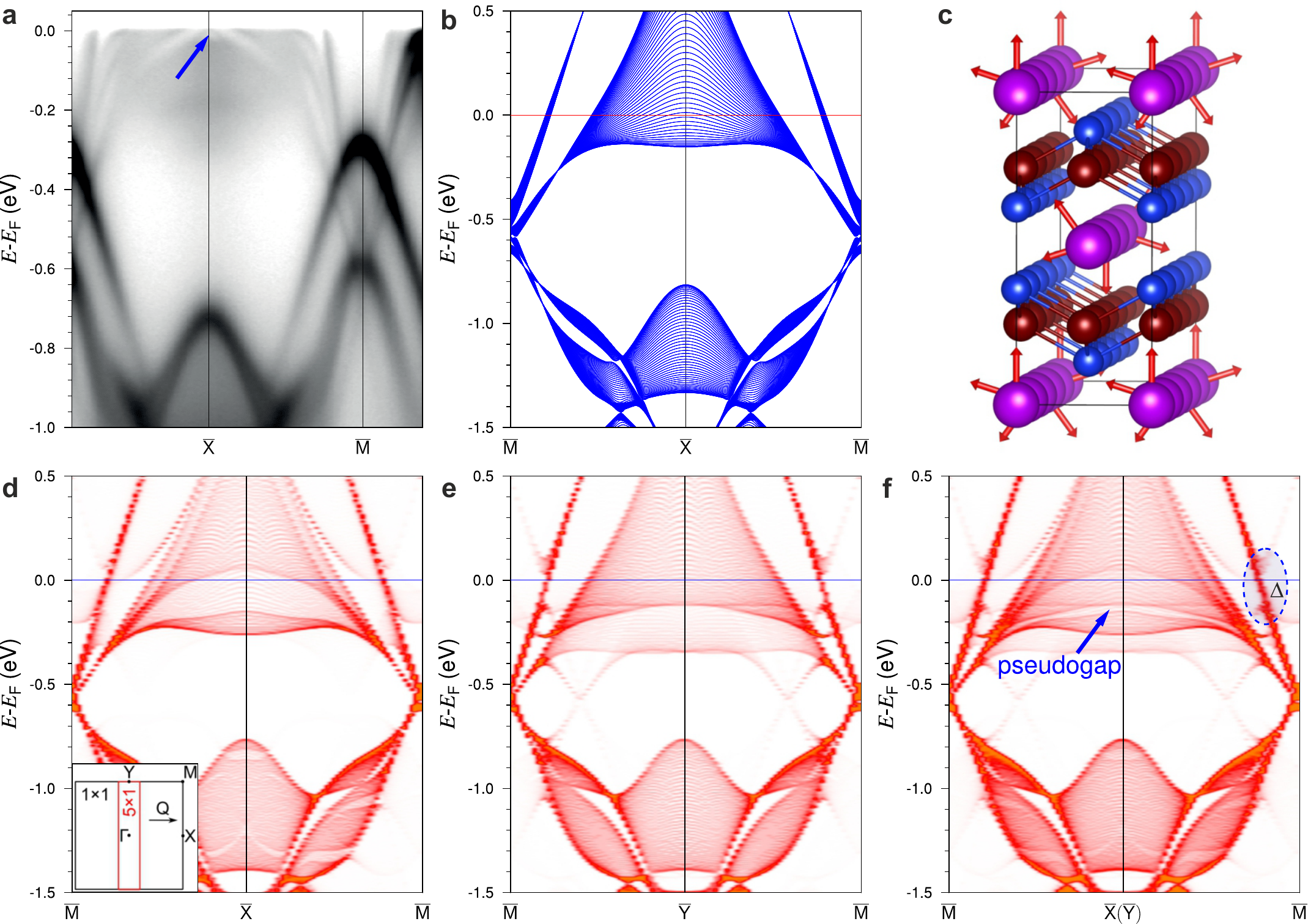}
  \caption{(a) Low-temperature ARPES data taken at 19~K along the $\overline{\mathrm{M}}-\overline{\mathrm{X}}-\overline{\mathrm{M}}$ direction of the surface BZ in the AFM phase of GdRu$_2$Si$_2$. A blue arrow indicates the ``sickle-moon" shaped pseudogap which appears in the bulk continuum states below $T_\mathrm{N}$. (b) Bulk electron states for the paramagnetic GdRu$_2$Si$_2$ computed along the $\overline{\mathrm{M}}-\overline{\mathrm{X}}-\overline{\mathrm{M}}$ direction and projected along $k_z$. (c) The $5 \times 1 \times 1$ supercell of GdRu$_2$Si$_2$ in the ordered phase where the Gd 4$f$ moments are rotated within the $yz$-plane with the propagation vector of $Q=0.2$ (in units $2\pi/a$), along the $x$ direction.  (d) The bulk-projected electron structure of the spiral-AFM ordered GdRu$_2$Si$_2$ calculated within the $5 \times 1$ supercell with the spin spiral propagated along $x$ unfolded onto the $\overline{\mathrm{M}}-\overline{\mathrm{X}}-\overline{\mathrm{M}}$ direction of the original $1 \times 1$ BZ. The inset shows the $1 \times 1$ (black) and $5 \times 1$ (red) BZs projected on (001). (e) The same as in panel (d) but for $\overline{\mathrm{M}}-\overline{\mathrm{Y}}-\overline{\mathrm{M}}$ direction. (f) Superposition of the bulk projected electron states shown in panels (d) and (e). The meaning of the highlighted $\Delta$ will be discussed later (see Fig.~\ref{fig6}d,e).
  }
  \label{fig3}
\end{figure*}

After cooling of the GdRu$_2$Si$_2$ sample below the spiral-AFM
phase transition to a temperature of \SI{19}{K}, we found that the
ARPES patterns did not reveal large modifications of the electronic
structure. Nevertheless, a number of subtle changes could be
observed. One of the most visible changes is the emergence of a
narrow bright feature (``pseudogap") in the bulk-projected spectrum,
which can be seen as a kind of ``sickle-moon" feature near
$E_\mathrm{F}$. This feature is presented in Fig.~\ref{fig3}a in the
ARPES data taken along the
$\overline{\mathrm{M}}-\overline{\mathrm{X}}-\overline{\mathrm{M}}$
direction of the surface BZ with the ``pseudogap" being indicated by
a blue arrow. For comparison, in Fig.~\ref{fig3}b we show the
projection of the calculated bulk states for the paramagnetic phase
onto the same $\overline{\mathrm{M}}
-\overline{\mathrm{X}}-\overline{\mathrm{M}}$ direction of the
surface BZ. It clearly demonstrates a homogeneous density of bulk
states near the $\overline{\mathrm{X}}$ point in the discussed
vicinity of the Fermi level. Having observed the ARPES-derived
signature of the spiral-AFM order of the Gd 4$f$ moments, our next
step aim was a corresponding appropriate theoretical modeling. Being
able to compute and achieve good agreement with experiment would
allow us to understand the ground state on the basis of first
principles calculations the reasons for such a peculiar magnetic
order in the AFM phase of GdRu$_2$Si$_2$.

To simulate the spiral-AFM phase, we constructed a bulk $5 \times 1
\times 1$ supercell, in which the spin spiral corresponding to
$Q=0.20$ (in units $2\pi/a$) is defined as schematically shown in
Fig.~\ref{fig3}c. In this supercell the spin magnetic moment on each
next Gd plane along the $x$ direction is rotated by $\phi=\ang{40}$.
Note that it is slightly different from $\phi=\ang{36}$ which
corresponds to the experimentally-derived incommensurate spiral
propagation vector of $Q=0.22$ ($2\pi/a$) \cite{Khanh_NatNano2020}.
As we will see below the small difference of $Q=0.20$ from $Q=0.22$
is not essential for the following analysis. The $k_z$-projected
bulk band structure unfolded onto the
$\overline{\mathrm{M}}-\overline{\mathrm{X}}-\overline{\mathrm{M}}$
direction (perpendicular to the spin spiral propagation,
Fig.~\ref{fig3}d) of the original $1 \times 1$ BZ shows that the
main changes compared to the paramagnetic case occur close to the
$\overline{\mathrm{X}}$ point and near $E_\mathrm{F}$ where density
of bulk states becomes inhomogeneous, having an increased density at
the very bottom and a reduced one at $E_\mathrm{F}$. The spectrum
calculated along the
$\overline{\mathrm{M}}-\overline{\mathrm{Y}}-\overline{\mathrm{M}}$
direction (along the spin spiral propagation, Fig.~\ref{fig3}e) also
demonstrates its main changes at $\overline{\mathrm{Y}}$ (which is
equivalent to $\overline{\mathrm{X}}$ in the original $1 \times 1$
BZ in the paramagnetic case) however, the reduced density can be
found here at lower energies while the increased density lies at
higher energies closer to $E_{\mathrm{F}}$. Taking into account that
the magnetic spirals along $x$ and $y$ directions must be
equiprobable due to symmetry reasons, and that ARPES should acquire
a mixture of magnetic domains with different spiral directions, the
superposition of the spectra calculated for directions along and
perpendicular to the magnetic spiral is presented in
Fig.~\ref{fig3}f, which shows the presence of a pseudogap in the
continuum of the bulk states below $E_\mathrm{F}$ (a narrow area of
a reduced density of the bulk states). This is in excellent
agreement with the experiment. Thus, the observation serves as a
direct photoemission evidence for the presence of the spiral-AFM
phase in the low-temperature limit.

\subsection{Electronic structure of the Si-terminated surface of the spiral-AFM GdRu$_2$Si$_2$}

\begin{figure*}
  \includegraphics[width=\textwidth]{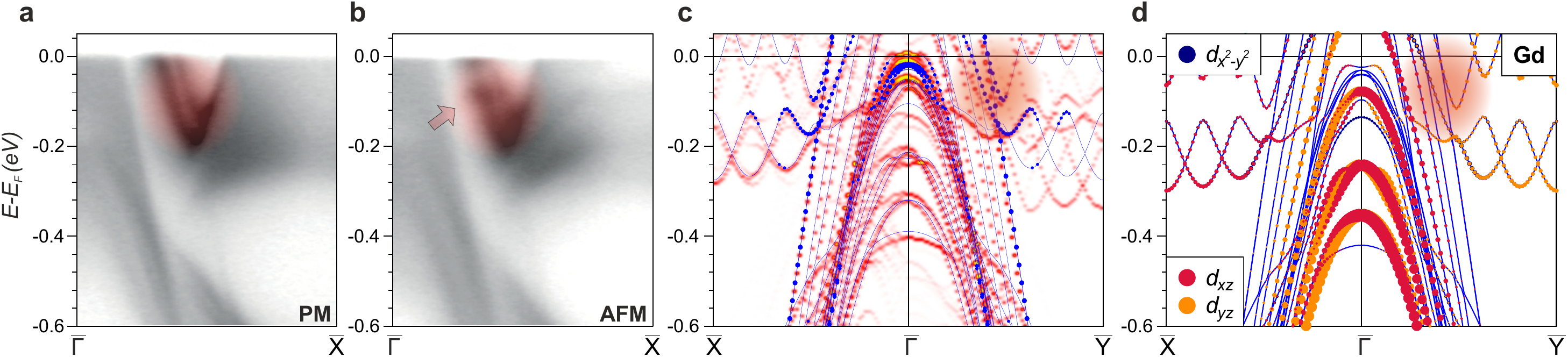}
  \caption{ARPES data taken along the $\overline{\Gamma}-\overline{\mathrm{X}}$ direction at \SI{50}{K} (a) and at \SI{19}{K} (b), that is, above and below the bulk $T_N$, respectively. The measurements were performed using a photon energy of \SI{50}{eV} with linear vertical ($s$) polarization. The orange highlighted spot marks an area of the surface resonant state. (c) Unfolded calculated spectrum (shown in red) of the Si-terminated GdRu$_2$Si$_2$ (001) surface for the spiral-AFM phase along $\overline{\Gamma}-\overline{\mathrm{X}}$ (along the propagation of the spin spiral) and $\overline{\Gamma}-\overline{\mathrm{Y}}$ (perpendicular to the spin spiral) directions. The spectrum of the GdRu$_2$Si$_2$ slab of the same thickness for the paramagnetic phase is imposed (blue curves) and weights of the surface states demonstrated by blue dots. (d) Weights of the Gd-5$d$ orbitals for the paramagnetic phase.
  }
 \label{fig4}
\end{figure*}

Next we explore how the spin-spiral AFM order is reflected in the
properties of the surface electrons states, in particular for the
Si-terminated surface of GdRu$_2$Si$_2$. Fig.~\ref{fig4}a and
Fig.~\ref{fig4}b show the ARPES data taken along the
$\overline{\Gamma}-\overline{\mathrm{X}}$ direction in the
paramagnetic and spiral-AFM phase, respectively.

The spectra reveal prominent bands located in the energy range of
\SI{-0.2}{eV} up to and crossing $E_\mathrm{F}$. Our calculations
allow us to identify these bands as surface resonances, which
penetrate deep into the material overlapping with the bulk band
states (Fig.~\ref{fig1}b and c, marked in green). Below the
spiral-AFM transition these states exhibit notable changes.
Particularly, in the paramagnetic phase the photoemission intensity
of these states varies smoothly along the spectral structure, while
at low temperature a oscillatory modification of the intensity
distribution can be observed. To reveal the origin of this intensity
redistribution, we expand the $5\times 1$ supercell bulk calculation
to a supercell slab calculation for the spiral-AFM phase.
Fig.~\ref{fig4}c shows the spectrum of the Si-terminated (001)
surface of GdRu$_2$Si$_2$ unfolded onto the
$\overline{\Gamma}-\overline{\mathrm{X}}$ (along the propagation of
the spin spiral) and the $\overline{\Gamma}-\overline{\mathrm{Y}}$
(perpendicular to the spin spiral) directions of the $1 \times 1$
BZ. Compared to the band structure of the paramagnetic phase (blue
lines in Fig.~\ref{fig4}c), both electronic structures almost
coincide for the direction of the spiral propagation
($\overline{\Gamma}-\overline{\mathrm{X}}$), whereas in the
perpendicular direction ($\overline{\Gamma}-\overline{\mathrm{Y}}$)
the band distribution considerably changes.

Near $E_\mathrm{F}$, the Gd-5$d$ contribution to these states is
almost equally determined by $d_{x^2-y^2}$ and $d_{xz}$($d_{yz}$)
orbitals (see the Fig.~\ref{fig4}d). With the spin spiral
propagating along the $x$ direction, the spin rotates in the
$yz$-plane and clearly affects the $d_{yz}$ orbitals but seems to
have no influence on the $d_{xz}$ orbital. Hence, when the surface
resonant state penetrates into these bulk states, where the weight
of Gd-5$d$ orbitals is considerable, it should experience the effect
of spiral-AFM order in the bulk. Again, due to the mixture of
$x$-and $y$-spiral magnetic domains, in ARPES this modification
manifests itself in both
$\text{-}\overline{\mathrm{X}}-\overline{\Gamma}$ and
$\overline{\Gamma}-\overline{\mathrm{X}}$ directions leading to a
subtle but clearly notable effect.

\subsection{Fermi surface and possible origin of the spin spiral}

\begin{figure*}
  \includegraphics[width=\textwidth]{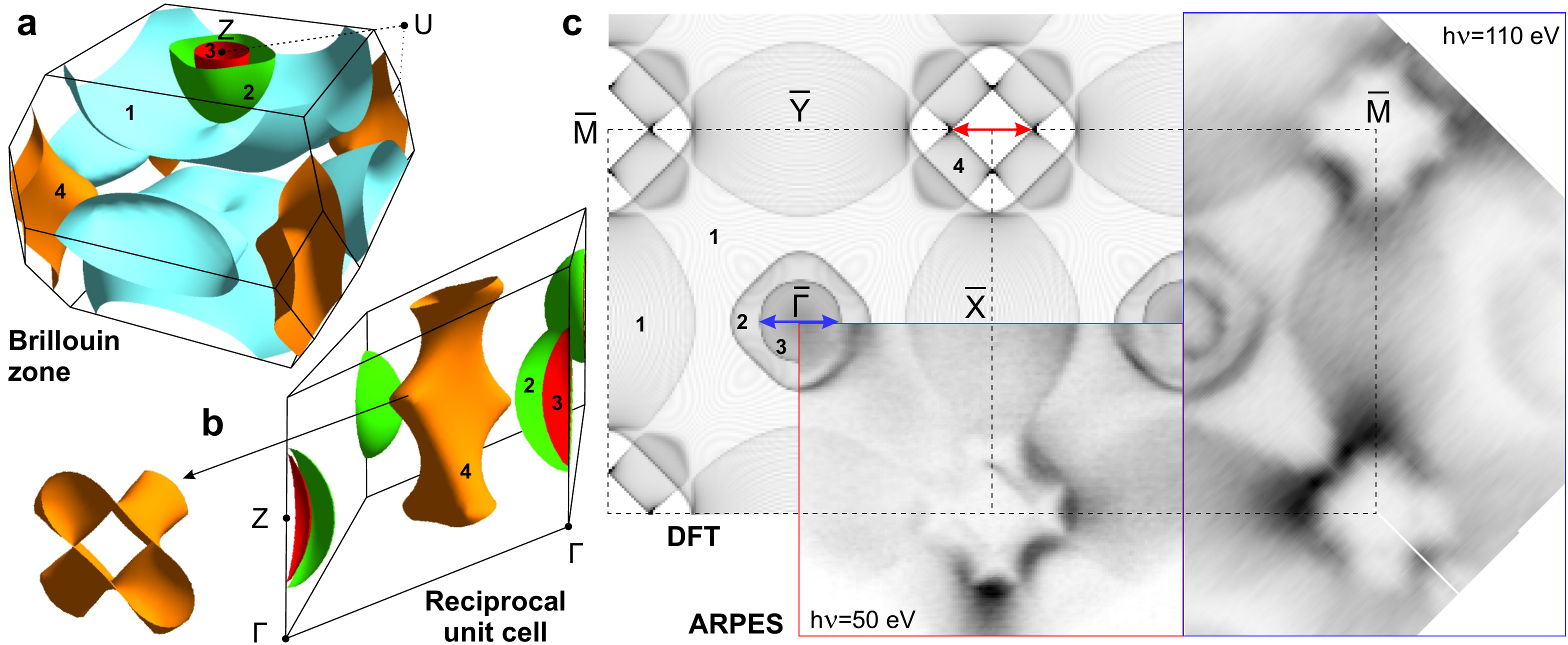}
  \caption{Fermi surface (FS) for GdRu$_2$Si$_2$ in the paramagnetic phase presented for the first BZ (a) and for the reciprocal lattice unit (b). In panel (b), band 1 is not shown for simplicity; the left outset shows the top perspective view of band 4. (c) Projection of the calculated FS on the (001) surface. The red and blue framed parts show the ARPES-derived FS maps taken with photon energies of $h\nu=\SI{50}{eV}$ and $h\nu=\SI{110}{eV}$, respectively. The red arrow indicates the proposed nested region in the 3D FS while the blue arrow shows the region near the center of the BZ, which was discussed in Ref.~\citenum{Bouaziz_PRL2022}.
  }
  \label{fig5}
\end{figure*}

In Ref.~\citenum{Bouaziz_PRL2022}, the properties of the
three-dimensional FS were intensively discussed. There, it was
argued that nesting properties found in the center of the BZ are the
key points for the formation of the spin-spiral magnetic order in
the AFM phase of GdRu$_2$Si$_2$ and define the skyrmion physics in
this material. To explore this point in detail, we present the
calculated FS for the paramagnetic phase of GdRu$_2$Si$_2$
(Fig.~\ref{fig5}a and b) with the experimentally-derived FS from
ARPES measurements (part of Fig.~\ref{fig5}c). To simplify the
analysis, we combine the ARPES data with the calculated band
structure projected along the $k_z$ direction onto the (001) surface
as shown on the left side of Fig.~\ref{fig5}c. Based on this
comparison, we see a close to perfect agreement between experiment
and theory, indicating that our calculations reliably describe all
bands in the vicinity of $E_\mathrm{F}$. The four bands which form
the FS are labeled as 1--4.

In Ref.~\citenum{Bouaziz_PRL2022} it was proposed that band 3, which
is highlighted in red in Fig.~\ref{fig5}a,b, is mostly barrel-shaped
and therefore provides good nesting conditions where the diameter of
the barrel fits well to the wave number $Q=0.22$ ($2\pi/a$) of the
spin spiral. We indicate the respective region by a blue arrow
around the $\overline{\Gamma}$ point in  Fig.~\ref{fig5}c. However,
based on our DFT results, we offer a different interpretation.
First, we find that band 3 differs considerably more from a
barrel-like shape than previously proposed and instead looks more
ellipsoidal with a significant $k_z$ dispersion centered around the
Z-point of the BZ. In this regard, our calculation suggest that
there are no strong nesting properties for this band.

On the other hand, our calculations indicate that band 4, which is
situated around the $\overline{\mathrm{M}}$ point, reveals nesting
properties. This band shows almost no $k_z$-dispersion in the
$\overline{\mathrm{M}}-\overline{\mathrm{X}}$ lateral direction. To
better visualize this fact, we present in the outset of
Fig.~\ref{fig5}b the discussed band 4 from the top perspective. In
this top view projection, it is clearly visible that the
$\overline{\mathrm{M}}-\overline{\mathrm{X}}$ cuts have no $k_z$
dispersion since they meet in single points. This nesting vector is
highlighted by a red arrow at the corner of BZ in the FS shown in
Fig.~\ref{fig5}c. Further, it is important to note that these points
have a high density of states and possess a strong contribution from
Gd 5$d$ states, which are essential for the exchange interaction and
will be discussed next.

\begin{figure*}
  \includegraphics[width=\linewidth]{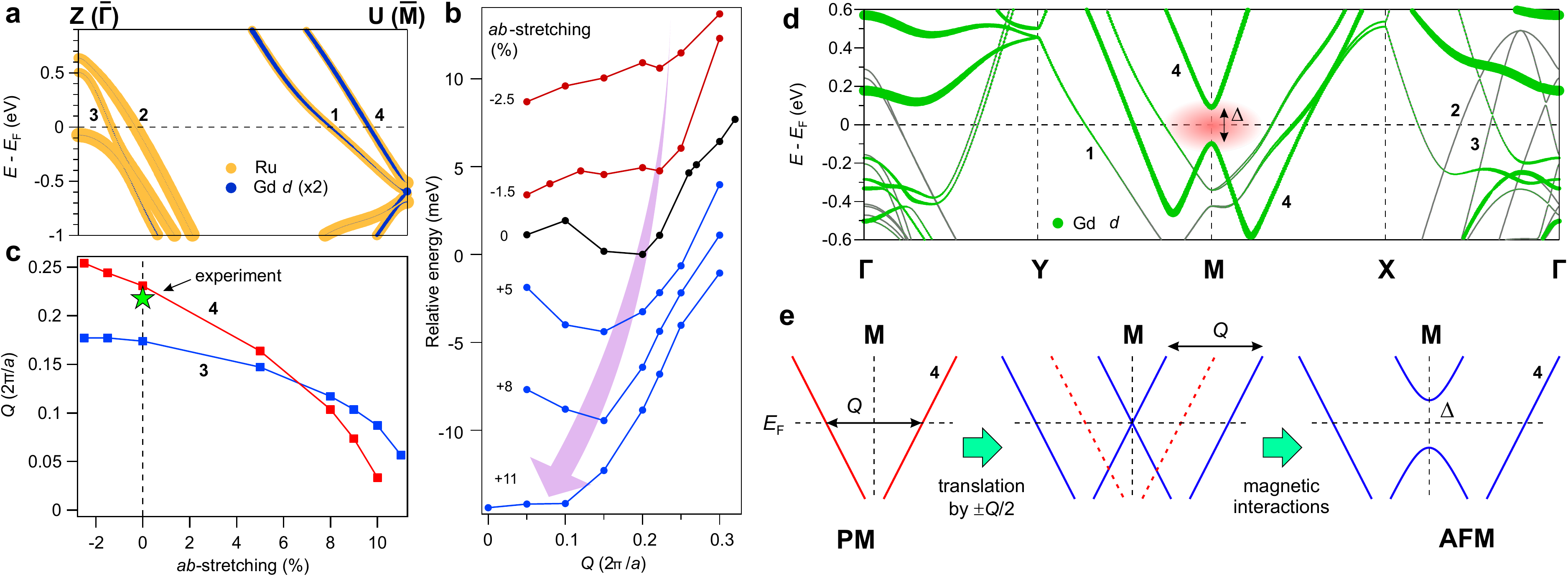}
  \caption{(a) Electron band structure near the $E_\mathrm{F}$ calculated along the Z--U direction of the BZ (see Fig.~\ref{fig5}a); the size of the colored symbols represents the weight of all states of Ru and 5$d$ states of Gd, where the weight for the latter is multiplied by a factor of two. (b) The evolution of the total energy of the tetragonal cell as a function of the spin-spiral wave number $Q$. The cartoon arrow indicates how the minimum of total energy evolves upon the lattice stretching. (c) Calculated evolution of the lengths of the red and blue $\vec{k}$ vectors shown in Fig.~\ref{fig5}c for the bands 3 and 4 at $E_\mathrm{F}$. (d) Band structure for the spiral-AFM state with $Q=0.22$ ($2\pi/a$), calculated using the tetragonal unit cell (the spiral vector $\vec{Q}$ is parallel to the $\Gamma-\mathrm{X}$ and $\mathrm{Y} - \mathrm{X}$ directions). (e) Schematic illustration of the formation of the band 4 in the direction along the spiral.
  }
 \label{fig6}
\end{figure*}

To understand the relation between the spiral period and the
features of the band structure, we analyze the orbital composition
of the bands 1--4 at $E_\mathrm{F}$. As illustrated in
Fig.~\ref{fig6}a, bands~1 and especially 4 have strong contributions
from Gd 5$d$ states, while bands 2 and 3 are mainly composed of Ru
states with only a small contribution from Gd states (see detailed
analysis of the orbital composition in the Appendix, Section 1).
This allows us to conclude that band 4 not only fulfills the nesting
condition, but also has the most significant admixture from Gd
5$d$-states, and thus is the most likely origin for the peculiar
magnetic structure of GdRu$_2$Si$_2$.

Next, we will test the relation between the nesting vector and the
spiral period. For that, we performed a series of ab-initio
calculations where the crystal lattice was stretched or compressed
within the $ab$ plane. Such deformations may lead to significant
changes of the band structure as well as the spiral period,
providing a possibility to explore the relationship between them.
Fig.~\ref{fig6}b demonstrates how the total energy evolves as a
function of the spiral wave number $Q$ for different values of
stretching. Note that the shown curves are arbitrarily shifted in
the energy scale for better comparison. For the case of the
equilibrium lattice (black curve), the energy minimum is located at
$Q=0.20$, which is very close to the experimental value of $Q=0.22$.
Upon expansion of the lattice the minimum shifts to lower $Q$ values
until the system becomes ferromagnetic which happens at
\SI{11}{\percent} of stretching. Upon compression of the lattice the
global energy minimum turns to a local minimum, which shifts towards
higher $Q$ values and almost disappears already at
\SI{2.5}{\percent} of compression.

We can now determine how the band structure evolves under lattice
deformation. For this, we will look in detail at the two most
interesting bands, namely band 4, which we propose to be responsible
for the spiral-AFM order since it exhibits strong nesting
properties, and band 3, which was proposed in
Ref.~\citenum{Bouaziz_PRL2022}. What we observe as a result of
stretching is that band 4 shifts up in energy, while band 3 gets
shifted down to lower energies. In both cases the related nesting
vector becomes shorter, which is illustrated in Fig.~\ref{fig6}c
where the lengths of the proposed nesting vectors are shown as
function of stretching. The nesting vectors were measured along the
Z--U direction of the BZ shown in Fig.~\ref{fig5}a. Our results
indicate a qualitative correlation between the length of the nesting
vectors and the wave number of the spiral. Upon stretching, the
nesting vector gets shorter for both bands 3 and 4, while in the
same way the spiral wave number becomes smaller. However, the
experimental value of the wave number $Q=0.22$ ($2\pi/a$) for the
equilibrium lattice  fits better to the nesting vector of band 4
than band 3 (experimental value shown as green star in
Fig.~\ref{fig6}c). Additionally, the stretching value at which band
4 shifts above $E_\mathrm{F}$ coincides with the stretching value of
\SI{11}{\percent} where the calculated $Q$ becomes zero. Thus, in
agreement with our previous statements about band 4, this band seems
to be the probable candidate for the  formation of the spin spiral
via $d-f$ hybridization.

To understand further the role of band 4 in the stabilization of the
spiral state, we present the theoretical band structure of the
spiral-AFM state in Fig.~\ref{fig6}d. These calculations were
performed using a so-called generalized Bloch condition. In this
formalism, for the case of a rotating spin, the wave function has
the form of the following spinor
\begin{equation}
  \Psi(\vec{r}) = \frac{1}{\sqrt{2}}\sum_{\vec{R}} \left(
  \begin{array}{c}
    {\rm e}^{i(\vec{k}-\vec{Q}/2)\cdot\vec{R}} \phi^{\uparrow}(\vec{r}-\vec{R}) \\
    {\rm e}^{i(\vec{k}+\vec{Q}/2)\cdot\vec{R}} \phi^{\downarrow}(\vec{r}-\vec{R})
  \end{array}
  \right) \,,
\end{equation}
summing over all unit cells. To ensure a spin rotation, the
quasi-momentum $\vec{k}$ gains an additional quantity of
$\pm\vec{Q}/2$ that results in the respective translation of bands
in the reciprocal space. This is schematically illustrated in
Fig.~\ref{fig6}e. In the paramagnetic phase, the states of band 4 at
$E_\mathrm{F}$ at the two sides of the M point are separated by the
nesting vector $\vec{Q}$ which coincides with the spin spiral wave
vector. Upon translation by $\pm\vec{Q}/2$ the two sides of the band
cross at the M point. The exchange interaction results in opening of
a magnetic gap $\Delta$ at the M point. The related decrease in the
energy of the occupied states results in stabilization of the
spiral. At the same time, bands 2 and 3 remain almost unchanged due
to their small contribution of the Gd 5$d$ orbitals.

It is worth noting that an experimental observation of the predicted
magnetic gap with ARPES is very challenging. First, the band
structure in Fig.~\ref{fig6}d cannot be directly compared with ARPES
data because of the redefinition of the quasimomentum $\vec{k}$. The
experimental measurements have to be compared with the unfolded
bands from the supercell calculations shown in Fig~\ref{fig3}f. One
can see that the discussed gap $\Delta$ in the band 4 is present in
the supercell calculation, although it appears shifted by
$\vec{Q}/2$ from the $\overline{\mathrm{M}}$ point. Moreover, it
does not look like a real gap since it is formed only between the
low-intensity replica bands which cannot be reliably detected in our
ARPES data. Instead, we were able to detect the aforementioned
pseudogap formed in band 1 that is of magnetic origin and thus
indicates the formation of the spiral-AFM state.

\subsection{Magnetic interaction}

To understand in more details the nature of magnetic interaction in
the system, we calculated exchange coupling constants from
first-principles applying the magnetic force theorem as it is
implemented within the multiple scattering theory
\cite{Liechtenstein1987,Hoffmann2020}. The exchange coupling
constants, $J_{ij}$ represent the direct and indirect overlap
between wave functions participating in the magnetic interaction
between atoms $i$ and $j$. A full Fourier transform, $J(\vec{Q})$ is
proportional the total energy, which minimum, i.e. the ground state,
corresponds to the maximum of $J(\vec{Q})$.  To determine the ground
state, $J(\vec{Q})$ was computed along high symmetry directions of
the BZ using various density functional approximations. The HSE06
hybrid functional is the best approach for describing the the
magnetic bulk band structure of GdRu$_2$Si$_2$ (see discussion in
the Appendix, Section 1). Unfortunately, this method is not yet
implemented within the multiple scattering theory. Therefore, we
used a GGA+$U$ functional to obtain exchange coupling constants. It
should be noted that previously reported first-principles studies of
this system were carried out using a bare GGA functional
\cite{Nomoto_PRL2020,Bouaziz_PRL2022}. Unfortunately, the GGA+$U$
approximation can not fully reproduce the band structure obtained
with the HSE06 hybrid functional. Nevertheless, the Gd $4f$ states
are pushed down in energy within this approaches and the band
structure in the Fermi level vicinity with exception several
features looks similar to that obtained within the HSE06  (Appendix,
Section 1, Fig.~\ref{fig_SI-1}).

 $J(\vec{Q})$ calculated along the high symmetry directions show three peaks: one is between $\Gamma$ and $\rm X$, the second is between $\rm X$ and $\rm M$, and the third is between $\Gamma$ and $\rm Z$ (see the Appendix, Section 2, Fig.~\ref{fig_SI-2}a). The global maximum, which corresponds the total energy minimum,  determines a spiral magnetic structure within the Gd layer. The $Q$ vector of this maximum is in a good agreement with our calculations within generalized Bloch conditions and experiment. However the presence of two other peaks with similar magnitudes indicates a competition between three various magnetic orders: two intra- and one inter-layer spirals. We point here that the results might be not fully correct, since the used approximation does not provide a full agreement with the HSE06 hybrid functional calculations. Another approach we have tried, a self-interaction correction (SIC) method, delivers a similar picture (see the Appendix, Fig.~\ref{fig_SI-2}b), although  $J(\vec{Q})$ are about two times smaller than that obtained within the GGA+$U$ approach. The later fact is because the SIC method overestimates usually localization of SI-corrected orbitals. However, both approaches predict a similar magnetic structure.

The next question arises of the origin of spirals in GdRu$_2$Si$_2$.
A structurally similar system, GdRh$_2$Si$_2$, exhibits a robust
layerwise antiferromagnetic ordering with the relatively high N\'eel
temperature of 107 K \cite{Guttler2016}. We think there are two main
reasons for so strong difference between these two AFM materials.
First, Rh has one $4d$ electron more than Ru. Therefore, Rh-Si
hybridization is essentially distinct from that between Ru and Si
atoms. The additional Rh $4d$ electron weakens the Gd-Si coupling
and Gd has more $5d$ electrons for the intra-layer magnetic coupling
between the Gd $4f$ local moments, which results in a strong
ferromagnetic order within the Gd layers. Indeed, the magnetic
interaction between the nearest Gd moments in GdRh$_2$Si$_2$ is much
stronger than that in GdRu$_2$Si$_2$: 5.1 meV vs. 0.7 meV, while the
inter-layer magnetic coupling in both system is of a similar
magnitude. The second reason for the stronger intra-layer coupling
in  GdRh$_2$Si$_2$ is the slight difference in the lattice
parameters in these materials. Lanthanides are well known for a
sensitive dependence of their magnetic properties on the structure.
Thus, a strong reduction of the intra-layer magentic interaction in
GdRu$_2$Si$_2$ leads to magnetic frustration and spiral formation in
this material. Our analysis indicates that the magnetic interaction
between Gd $4f$ local moments is mainly mediated by conduction
electrons. Although our ARPES experiments and DFT calculations do
not confirm the nesting found in Ref.~\cite{Bouaziz_PRL2022}, the
magnitude of the respective vector which was discussed in
Ref.~\cite{Bouaziz_PRL2022} is similar to the nesting vector at the
corner of BZ which was found in our study. Combining our finding and
the discussion made in Ref.~\cite{Bouaziz_PRL2022} we conclude that
the RKKY interaction is the dominant mechanism for helical magnetism
in this system.

\section{Conclusions}

In summary, applying momentum-resolved photoemission measurements
and \emph{ab-initio} DFT calculations, we explored the bulk and
surface electronic structure of the helical antiferromagnetic
material GdRu$_2$Si$_2$. The ARPES-derived data, taken for
GdRu$_2$Si$_2$ in paramagnetic phase, reveal a sharp patterns
containing the surface and bulk-projected electron states which
allow to explore how these states are modified upon the PM-AFM phase
transition.  Note that in difference to many recently studied
RET$_2$Si$_2$ non-helical antiferromagnets \cite{Gustav2022}, the
surface states of Si-terminated GdRu$_2$Si$_2$ do not reveal notable
exchange splittings. The latter is mainly because these states are
localized on the Ru layer in Si-Ru-Si surface block. Comparison of
the ARPES data taken from the AFM phase of GdRu$_2$Si$_2$ with the
results of \emph{ab-initio} DFT calculations allowed us to find
bulk-related spectral features which are intrinsic for the spiral's
order of 4$f$ moments. Namely, we detected a pseudogap within the
bulk continuum states close to $E_\mathrm{F}$ which is seen as a
``sickle-moon" shaped spectral feature. A rather good agreement of
experimental and theoretical results allowed us to characterize in
detail the properties and orbital composition of the Fermi surface
of GdRu$_2$Si$_2$. Note that our results do not confirm the
prediction of the existence of the nested three-dimensional
barrel-shaped FS near the $\Gamma$-point which was reported recently
\cite{Bouaziz_PRL2022}. Instead, we found a nested FS sheet at the
corner of the BZ, supporting the recent results reported in
Ref.~\citenum{Matsuyama_PRB23}. Our theoretical analysis suggests
that this feature possesses a strong admixture of $5d$-states of Gd.
We show that stretching or compressing the crystal lattice within
the $ab$ plane essentially influences the Gd $5d$ states nesting
vector and consequently changes the period of the magnetic spiral in
the Gd $4f$ spin channel. The obtained results allow us to conclude
that the RKKY interaction is the most plausible mechanism which
defines the spiral magnetic order in GdRu$_2$Si$_2$ and is
responsible for the emergence of skyrmions in this material.

\section*{Methods}

Single crystals of GdRu$_2$Si$_2$ were grown from indium flux using
high purity starting materials Gd (99.9\%, EvoChem), Ru (99.95\%,
EvoChem), Si (99.9999\%, Wacker) and In (99.9995\%, Schuckard) and a
modified Bridgman method as described in \cite{Kliemt2019}. Gd, Ru,
Si, and In were used in the ratio of Gd : Ru : Si : In = 1 : 2 : 2 :
24. The crystal growth was performed in a vertical furnace (GERO
HTRV 70-250/18) (T$_{max}=1600^{\circ}$C), a slow cooling period
with a rate of $1-4\,\rm K/h$ down to $850^{\circ}$C followed by
fast cooling to room temperature with $300\,\rm K/h$. The crystals
were separated from the flux by etching in hydrochloric acid. We
obtained platelet-shaped crystals with typical dimensions of $3\,\rm
mm \times 3\,\rm mm$ and a thickness of $50-100\,\mu m$.

ARPES experiments were first performed at the ULTRA endstation of
the SIS-X09LA beamline, Swiss Light Source, in November 2021.
Additional detailed high-quality measurements were obtained in a
second measurement at the BLOCH beamline, MAX-IV laboratory, in July
2022. Both ARPES endstations were equipped with a Scienta R4000
analyzer. The single-crystal samples of GdRu$_2$Si$_2$ were cleaved
\emph{in situ} under ultra-high vacuum conditions better than
\SI{e-10}{mmar}.

Electronic structure calculations were carried out within the
density functional theory using the projector augmented-wave (PAW)
method \cite{Blochl.prb1994,PAW1,PAW2} as implemented in the VASP
code \cite{vasp1,vasp2}. The exchange-correlation energy was treated
using the generalized gradient approximation \cite{Perdew.prl1996}
for most calculations. The trivalent Gd potential (when strongly
localized valence 4$f$ electrons are treated as core states) was
used for non-magnetic calculations. The standard tetravalent Gd
potential in which the 4$f$ electrons are treated as valence states
was used for spin-polarized calculations of magnetic phases. To
correctly describe the highly correlated Gd-4$f$ electrons, we
include the correlation effects within both the HSE06 screened
hybrid functional \cite{HSE06} and the GGA+$U$
method~\cite{Anisimov1991}. The values of $U$ and $J$ parameters
were taken to be of 6.7~eV and 0.7~eV, respectively, which give a
good agreement with HSE06 band structure. Additionally we use
Slater-type DFT-1/2 self-energy correction method
\cite{DFT12_1,DFT12_2} with a partially (quaternary) ionized silicon
potential for better describing hybridization between deep Gd-$f$
and Si-$p_{xy}$ orbitals. To simulate the spiral magnetic state we
consider spin spirals with two different methods. The first is to
construct explicitly the spiral in supercell using noncollinear
magnetic moments in the $yz$-plane, perpendicular to the spin spiral
propagation vector (along $x$ direction) and the second approach is
to use a spin spiral method to simulate spirals in the unit cell
with generalized Bloch conditions \cite{Sandratskii1998}. In the
first case we use BandUP code
\cite{Unfolding-1-PRB-2014,Unfolding-2-PRB-2015} to unfold the
supercell band structure onto $1\times 1$ BZ. Additional band
structure calculations for the paramagnetic phase (Fig.~\ref{fig5}c)
were performed with the Gd 4$f$ states in the core using
FPLO-18.00-52 code (improved version of the original FPLO code by
K.~Koepernik and H.~Eschrig \cite{FPLO}). The results are in perfect
agreement with those obtained with the VASP code. All presented
ball-and-stick atomic structures were visualized with {\sc vesta}
\cite{VESTA}. The Fermi surface was determined on a dense $27 \times
27 \times 23$ $k$-point mesh and visualized by using
\texttt{FermiSurfer} \cite{Kawamura2019}.

\section*{Acknowledgements}
We acknowledge the German Research Foundation (DFG) for the support
through the grants No. KR3831/5-1, No. LA655/20-1, SFB1143 (Project
No. 247310070), and TRR288 (No. 422213477, Project No. A03. The
density functional theory calculations were supported by the
Government research assignment for ISPMS SB RAS (Project
FWRW-2022-0001). E.V.C. acknowledges support from Saint Petersburg
State University (Project ID No. 94031444). V.S.S. and D.Yu.U. work
was partially supported by the Ministry of Science and Higher
Education of the Russian Federation (No. FSMG-2023-0014) and RSF
23-72-30004. The calculations were partially performed using the
equipment of the Shared Resource Center ``Far Eastern Computing
Resource" of IACP FEB RAS (https://cc.dvo.ru) and Joint
Supercomputer Center of the Russian Academy of Sciences
(https://rscgroup.ru/en/project/jscc). We also thank the Paul
Scherrer Institut, Villigen, Switzerland for the allocation of ARPES
experiments at the ULTRA endstation of the SIS-X09LA beamline of the
Swiss Light Source. We acknowledge MAX IV Laboratory for
experimental time on Beamline BLOCH under Proposal 20211066.
Research conducted at MAX IV, a Swedish national user facility, is
supported by the Swedish Research Council under contract 2018-07152,
the Swedish Governmental Agency for Innovation Systems under
contract 2018-04969, and Formas under contract 2019-0249.

\appendix
\section*{Appendix}

\subsection{Calculated bulk band structure}

\begin{figure*}
\includegraphics[width=\textwidth]{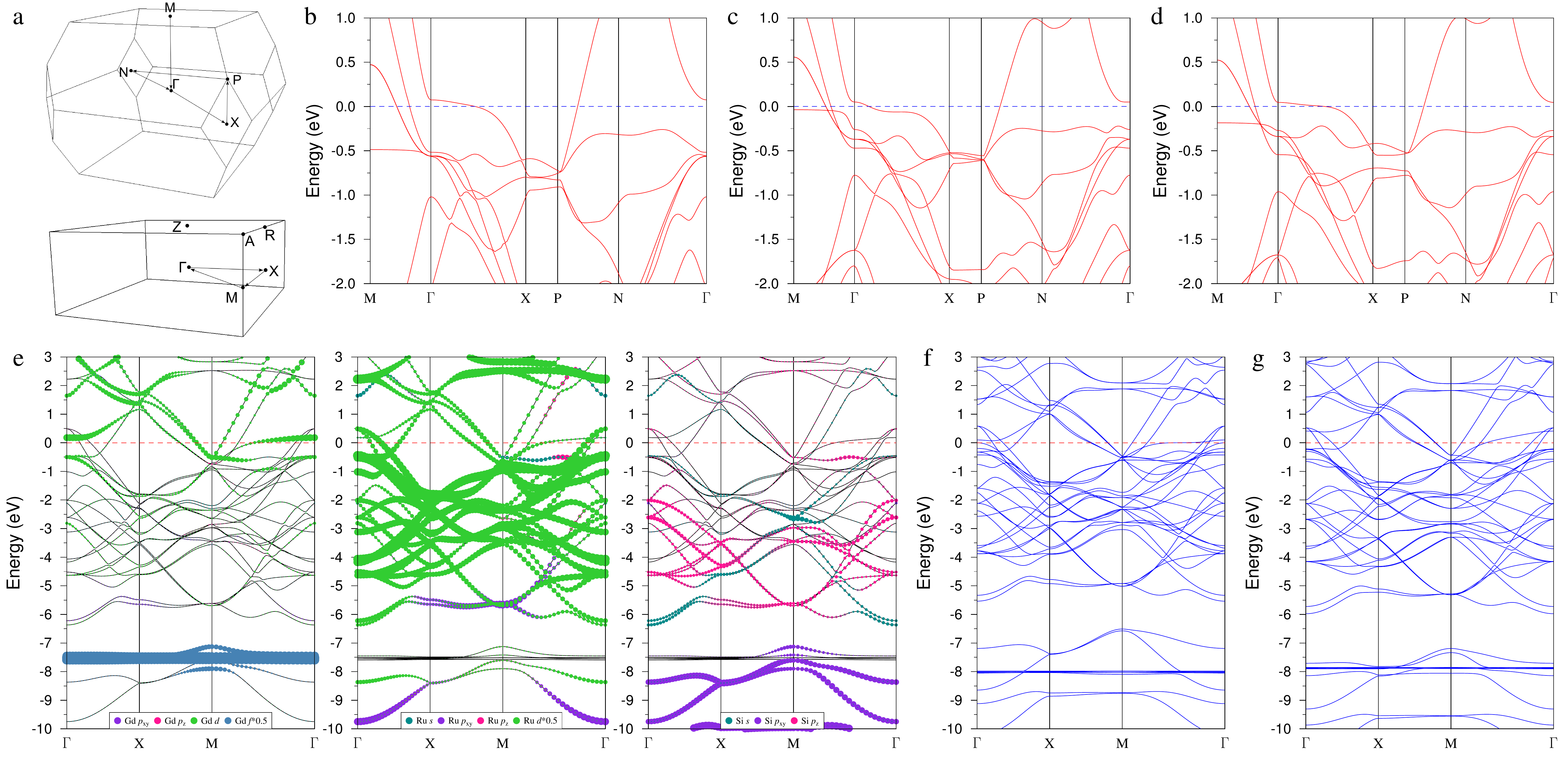}
\caption{ (a) Bulk Brillouin zones for paramagnetic primitive  (top)
and for AFM tetragonal (bottom) cells. Arrows show the paths along
which electronic spectra are calculated. (b) Bulk energy spectrum of
paramagnetic GdRu$_2$Si$_2$ where Gd states are treated as core
electrons as obtained with HSE06 hybrid functional. (c) The spectrum
of nonmagnetic GdRu$_2$Si$_2$ calculated within bare GGA-PBE
approach. (d) The same but within DFT-1/2 approach. (e)
Scalar-relativistic spectrum of AFM GdRu$_2$Si$_2$ calculated within
HSE06 approach with weights of the Gd (left), Ru (middle), and Si
(right) orbitals. (f) Scalar-relativistic spectrum of AFM phase
calculated within GGA-PBE with $U=6.7$ eV and $J=0.7$ eV. (g) The
spectrum calculated within GGA-PBE with $U=6.7$ eV and $J=0.7$ eV
and quaternary ionized Si $p$ states within DFT-1/2 approach.
 }
 \label{fig_SI-1}
\end{figure*}

To reveal the band structure of GdRu$_2$Si$_2$ with helical AFM
magnetic structure, we started from paramagnetic calculation where
trivalent Gd potential is used in which strongly localized valence
$4f$ electrons are treated as core states (Fig.~\ref{fig_SI-1}a).
The spectra of paramagnetic phase calculated along high symmetry
directions of the bulk Brillouin zone (Fig.~\ref{fig_SI-1}a, top)
using the Heyd-Scuseria-Ernzerhof (HSE06) screened hybrid functional
\cite{HSE06} and that obtained within bare GGA-PBE calculation are
demonstrated in Figs.~\ref{fig_SI-1} a and b, respectively. As can
be seen from comparison of these spectra the latter one does not
reproduce well the more accurate HSE06 result, especially at the
Fermi level. Applying DFT-1/2 corrections over GGA-PBE allows
receiving more accurate description of the nonmagnetic spectrum in
this energy range (Fig.~\ref{fig_SI-1}d). Scalar-relativistic
spectrum of AFM GdRu$_2$Si$_2$ was calculated with tetravalent Gd
potential in which the $4f$ electrons are treated as valence states
within the HSE06 approach (Fig.~\ref{fig_SI-1}e).
Fig.~\ref{fig_SI-1}e also presents projections of weights of the Gd,
Ru, and Si orbitals (left, middle, and right subpanels,
respectively). As can be seen the Gd $4f$ localized band, located at
$\sim -7.5$ eV, hybridizes with Si $p_{xy}$ orbitals. Si $s$ and
$p_z$ states contribute to the bottom of the main valence band,
which is mainly formed by Ru $d$ orbitals. However, the Gd $d$
states also present near the Fermi level ($E_{\rm F}$) where they
have most pronounced weights in the vicinity of the M point. Thus we
can conclude the Gd $d$ are the orbitals which mediate magnetization
of the bulk bands near the Fermi level via Gd $f-d$ hybridization.
To mimic the HSE06 band structure the simplified GGA+$U$ approach
was applied (see details in the Method section). The GGA+$U$
spectrum as obtained within PBE calculation reasonably well
reproduces the position of the Gd $f-d$ band (Fig.~\ref{fig_SI-1}f).
However, like in the nonmagnetic case, the spectrum at $E_{\rm F}$
differs from that calculated within HSE06 approach as well as
position of the deep Si $p_{xy}$ states. The partial ionization of
Si potential within DFT-1/2 method resolves both these problems
(Fig.~\ref{fig_SI-1}g).

\subsection{Magnetic interaction}

\begin{figure*}
\includegraphics[width=0.75\textwidth]{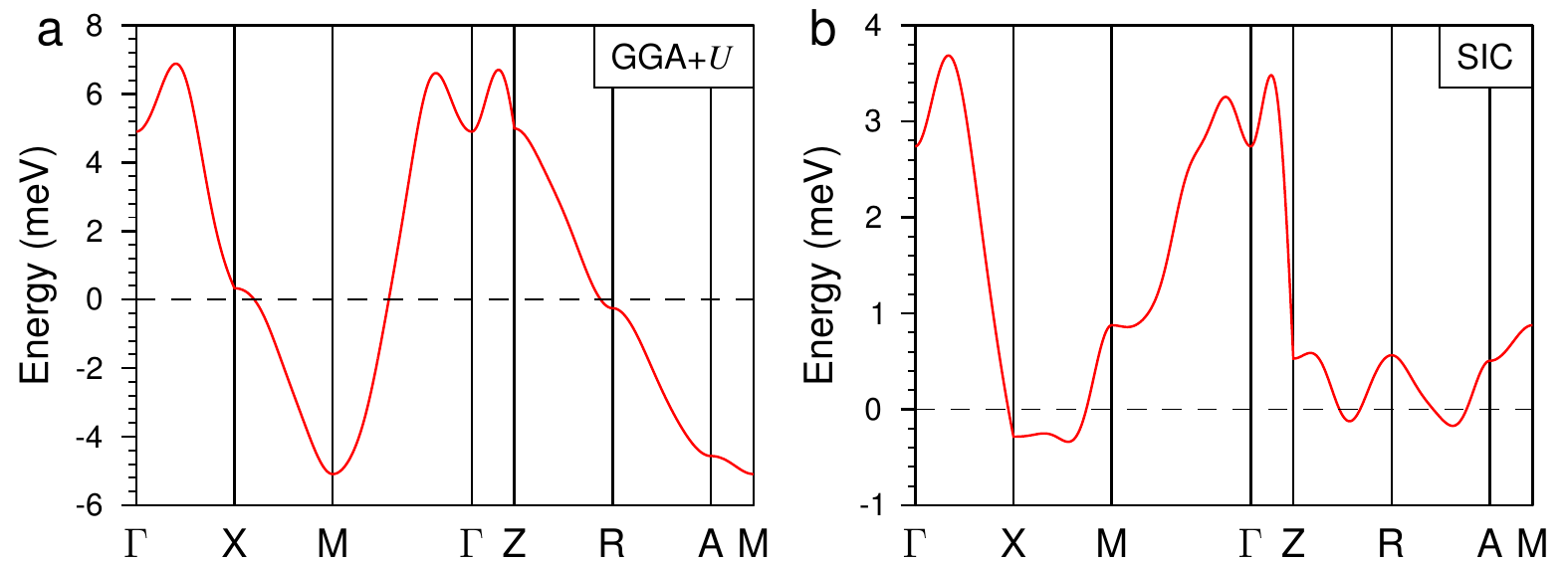}
\caption{$J(\vec{Q})$ calculated along the high symmetry directions
of the BZ of the tetragonal cell. The global maximum of $J(\vec{Q})$
corresponds to the total energy minimum. Results are presented for a
GGA+$U$ (a) and self-interaction correction (SIC) (b) DFT
functionals.}
 \label{fig_SI-2}
\end{figure*}

Exchange constants were calculated using the magnetic force theorem
as it is implemented within the multiple scattering
theory~\cite{Liechtenstein1987}. To describe localized Gd $4f$
electrons we used two different DFT functionals, a GGA+$U$
approach~\cite{Dudarev.prb1998} and a self-interaction correction
(SIC) method~\cite{Perdew1981,Luders2005,Hoffmann2020}. The results
of our calculations are presented in Fig.~\ref{fig_SI-2}. Both
approaches provide a similar behaviour  of $J(\vec{Q})$ and the
$J(\vec{Q})$ maximum occurs at the same wave vector
$Q\approx$0.2~($2\pi/a$) along $\Gamma-{\rm X}$ direction. However,
the magnitude of $J$'s calculated within the SIC method is
significantly smaller than that obtained using the GGA+$U$
approximation. The reason is that the SIC method overestimates
localization of corrected orbitals. Therefore, in our manuscript we
rely on the results obtained within the GGA+$U$ approach.

\begin{thebibliography}{38}%
\makeatletter
\providecommand \@ifxundefined [1]{%
 \@ifx{#1\undefined}
}%
\providecommand \@ifnum [1]{%
 \ifnum #1\expandafter \@firstoftwo
 \else \expandafter \@secondoftwo
 \fi
}%
\providecommand \@ifx [1]{%
 \ifx #1\expandafter \@firstoftwo
 \else \expandafter \@secondoftwo
 \fi
}%
\providecommand \natexlab [1]{#1}%
\providecommand \enquote  [1]{``#1''}%
\providecommand \bibnamefont  [1]{#1}%
\providecommand \bibfnamefont [1]{#1}%
\providecommand \citenamefont [1]{#1}%
\providecommand \href@noop [0]{\@secondoftwo}%
\providecommand \href [0]{\begingroup \@sanitize@url \@href}%
\providecommand \@href[1]{\@@startlink{#1}\@@href}%
\providecommand \@@href[1]{\endgroup#1\@@endlink}%
\providecommand \@sanitize@url [0]{\catcode `\\12\catcode `\$12\catcode
  `\&12\catcode `\#12\catcode `\^12\catcode `\_12\catcode `\%12\relax}%
\providecommand \@@startlink[1]{}%
\providecommand \@@endlink[0]{}%
\providecommand \url  [0]{\begingroup\@sanitize@url \@url }%
\providecommand \@url [1]{\endgroup\@href {#1}{\urlprefix }}%
\providecommand \urlprefix  [0]{URL }%
\providecommand \Eprint [0]{\href }%
\providecommand \doibase [0]{https://doi.org/}%
\providecommand \selectlanguage [0]{\@gobble}%
\providecommand \bibinfo  [0]{\@secondoftwo}%
\providecommand \bibfield  [0]{\@secondoftwo}%
\providecommand \translation [1]{[#1]}%
\providecommand \BibitemOpen [0]{}%
\providecommand \bibitemStop [0]{}%
\providecommand \bibitemNoStop [0]{.\EOS\space}%
\providecommand \EOS [0]{\spacefactor3000\relax}%
\providecommand \BibitemShut  [1]{\csname bibitem#1\endcsname}%
\let\auto@bib@innerbib\@empty
\bibitem [{\citenamefont {{\'{S}}laski}\ and\ \citenamefont
  {Szytu{\l}a}(1982)}]{Slaski_JLCM1982}%
  \BibitemOpen
  \bibfield  {author} {\bibinfo {author} {\bibfnamefont {M.}~\bibnamefont
  {{\'{S}}laski}}\ and\ \bibinfo {author} {\bibfnamefont {A.}~\bibnamefont
  {Szytu{\l}a}},\ }\bibfield  {title} {\bibinfo {title} {Crystal structure and
  magnetic properties of {REMe$_2$Si$_2$} compounds {(RE = Gd, Dy, Ho,
  Er; Me = Ru, Rh, Pd, Ir)}},\ }\href
  {https://doi.org/10.1016/0022-5088(82)90099-6} {\bibfield  {journal}
  {\bibinfo  {journal} {Journal of the Less-Common Metals}\ }\textbf {\bibinfo
  {volume} {87}},\ \bibinfo {pages} {L1 -- L3} (\bibinfo {year}
  {1982})}\BibitemShut {NoStop}%
\bibitem [{\citenamefont {Hiebl}\ \emph {et~al.}(1983)\citenamefont {Hiebl},
  \citenamefont {Horvath}, \citenamefont {Rogl},\ and\ \citenamefont
  {Sienko}}]{Hiebl_JMMM1983}%
  \BibitemOpen
  \bibfield  {author} {\bibinfo {author} {\bibfnamefont {K.}~\bibnamefont
  {Hiebl}}, \bibinfo {author} {\bibfnamefont {C.}~\bibnamefont {Horvath}},
  \bibinfo {author} {\bibfnamefont {P.}~\bibnamefont {Rogl}},\ and\ \bibinfo
  {author} {\bibfnamefont {M.~J.}\ \bibnamefont {Sienko}},\ }\bibfield  {title}
  {\bibinfo {title} {Magnetic properties and structural chemistry of ternary
  silicides {(RE,Th,U)Ru$_2$Si$_2$ (RE = Rare Earth)}},\ }\href
  {https://doi.org/10.1016/0304-8853(83)90058-6} {\bibfield  {journal}
  {\bibinfo  {journal} {Journal of Magnetism and Magnetic Materials}\ }\textbf
  {\bibinfo {volume} {37}},\ \bibinfo {pages} {287} (\bibinfo {year}
  {1983})}\BibitemShut {NoStop}%
\bibitem [{\citenamefont {{\'{S}}laski}\ \emph {et~al.}(1984)\citenamefont
  {{\'{S}}laski}, \citenamefont {Szytu{\l}a}, \citenamefont {Leciejewicz},\
  and\ \citenamefont {Zygmunt}}]{Slaski_JMMM1984}%
  \BibitemOpen
  \bibfield  {author} {\bibinfo {author} {\bibfnamefont {M.}~\bibnamefont
  {{\'{S}}laski}}, \bibinfo {author} {\bibfnamefont {A.}~\bibnamefont
  {Szytu{\l}a}}, \bibinfo {author} {\bibfnamefont {J.}~\bibnamefont
  {Leciejewicz}},\ and\ \bibinfo {author} {\bibfnamefont {A.}~\bibnamefont
  {Zygmunt}},\ }\bibfield  {title} {\bibinfo {title} {Magnetic properties of
  {RERu$_2$Si$_2$ (RE=Pr, Nd, Gd, Tb, Dy, Er)} intermetallics},\ }\href
  {https://www.sciencedirect.com/science/article/pii/0304885384903482}
  {\bibfield  {journal} {\bibinfo  {journal} {Journal of Magnetism and Magnetic
  Materials}\ }\textbf {\bibinfo {volume} {46}},\ \bibinfo {pages} {114}
  (\bibinfo {year} {1984})}\BibitemShut {NoStop}%
\bibitem [{\citenamefont {Khanh}\ \emph {et~al.}(2020)\citenamefont {Khanh},
  \citenamefont {Nakajima}, \citenamefont {Yu}, \citenamefont {Gao},
  \citenamefont {Shibata}, \citenamefont {Hirschberger}, \citenamefont
  {Yamasaki}, \citenamefont {Sagayama}, \citenamefont {Nakao}, \citenamefont
  {Peng}, \citenamefont {Nakajima}, \citenamefont {Takagi}, \citenamefont
  {Tokura},\ and\ \citenamefont {Seki}}]{Khanh_NatNano2020}%
  \BibitemOpen
  \bibfield  {author} {\bibinfo {author} {\bibfnamefont {N.~D.}\ \bibnamefont
  {Khanh}}, \bibinfo {author} {\bibfnamefont {T.}~\bibnamefont {Nakajima}},
  \bibinfo {author} {\bibfnamefont {X.}~\bibnamefont {Yu}}, \bibinfo {author}
  {\bibfnamefont {S.}~\bibnamefont {Gao}}, \bibinfo {author} {\bibfnamefont
  {K.}~\bibnamefont {Shibata}}, \bibinfo {author} {\bibfnamefont
  {M.}~\bibnamefont {Hirschberger}}, \bibinfo {author} {\bibfnamefont
  {Y.}~\bibnamefont {Yamasaki}}, \bibinfo {author} {\bibfnamefont
  {H.}~\bibnamefont {Sagayama}}, \bibinfo {author} {\bibfnamefont
  {H.}~\bibnamefont {Nakao}}, \bibinfo {author} {\bibfnamefont
  {L.}~\bibnamefont {Peng}}, \bibinfo {author} {\bibfnamefont {K.}~\bibnamefont
  {Nakajima}}, \bibinfo {author} {\bibfnamefont {T.}~\bibnamefont {Takagi},
  \bibfnamefont {R.~Arima}}, \bibinfo {author} {\bibfnamefont {Y.}~\bibnamefont
  {Tokura}},\ and\ \bibinfo {author} {\bibfnamefont {S.}~\bibnamefont {Seki}},\
  }\bibfield  {title} {\bibinfo {title} {Nanometric square skyrmion lattice in
  a centrosymmetric tetragonal magnet},\ }\href
  {https://doi.org/10.1038/s41565-020-0684-7} {\bibfield  {journal} {\bibinfo
  {journal} {Nature Nanotechnology}\ }\textbf {\bibinfo {volume} {15}},\
  \bibinfo {pages} {444--449} (\bibinfo {year} {2020})}\BibitemShut {NoStop}%
\bibitem [{\citenamefont {Garnier}\ \emph {et~al.}(1995)\citenamefont
  {Garnier}, \citenamefont {Gignoux}, \citenamefont {Iwata}, \citenamefont
  {Schmitt}, \citenamefont {Shigeoka},\ and\ \citenamefont
  {Zhang}}]{Garnier_JMMM1995}%
  \BibitemOpen
  \bibfield  {author} {\bibinfo {author} {\bibfnamefont {A.}~\bibnamefont
  {Garnier}}, \bibinfo {author} {\bibfnamefont {D.}~\bibnamefont {Gignoux}},
  \bibinfo {author} {\bibfnamefont {N.}~\bibnamefont {Iwata}}, \bibinfo
  {author} {\bibfnamefont {D.}~\bibnamefont {Schmitt}}, \bibinfo {author}
  {\bibfnamefont {T.}~\bibnamefont {Shigeoka}},\ and\ \bibinfo {author}
  {\bibfnamefont {F.}~\bibnamefont {Zhang}},\ }\bibfield  {title} {\bibinfo
  {title} {Anisotropic metamagnetism in {GdRu$_2$Si$_2$}},\ }\href
  {https://doi.org/10.1016/0304-8853(94)00783-7} {\bibfield  {journal}
  {\bibinfo  {journal} {Journal of Magnetism and Magnetic Materials}\ }\textbf
  {\bibinfo {volume} {140-144}},\ \bibinfo {pages} {899} (\bibinfo {year}
  {1995})}\BibitemShut {NoStop}%
\bibitem [{\citenamefont {Rotter}\ \emph {et~al.}(2007)\citenamefont {Rotter},
  \citenamefont {Doerr}, \citenamefont {Zschintzsch}, \citenamefont {Lindbaum},
  \citenamefont {Sassik},\ and\ \citenamefont {Behr}}]{Rotter_JMMM2007}%
  \BibitemOpen
  \bibfield  {author} {\bibinfo {author} {\bibfnamefont {M.}~\bibnamefont
  {Rotter}}, \bibinfo {author} {\bibfnamefont {M.}~\bibnamefont {Doerr}},
  \bibinfo {author} {\bibfnamefont {M.}~\bibnamefont {Zschintzsch}}, \bibinfo
  {author} {\bibfnamefont {A.}~\bibnamefont {Lindbaum}}, \bibinfo {author}
  {\bibfnamefont {H.}~\bibnamefont {Sassik}},\ and\ \bibinfo {author}
  {\bibfnamefont {G.}~\bibnamefont {Behr}},\ }\bibfield  {title} {\bibinfo
  {title} {The magnetoelastic paradox in {GdAg$_2$} and {GdRu$_2$Si$_2$}},\
  }\href {https://doi.org/10.1016/j.jmmm.2006.10.401} {\bibfield  {journal}
  {\bibinfo  {journal} {Journal of Magnetism and Magnetic Materials}\ }\textbf
  {\bibinfo {volume} {310}},\ \bibinfo {pages} {1383} (\bibinfo {year}
  {2007})}\BibitemShut {NoStop}%
\bibitem [{\citenamefont {Samanta}\ \emph {et~al.}(2008)\citenamefont
  {Samanta}, \citenamefont {Das},\ and\ \citenamefont
  {Banerjee}}]{Samanta_JAP2008}%
  \BibitemOpen
  \bibfield  {author} {\bibinfo {author} {\bibfnamefont {T.}~\bibnamefont
  {Samanta}}, \bibinfo {author} {\bibfnamefont {I.}~\bibnamefont {Das}},\ and\
  \bibinfo {author} {\bibfnamefont {S.}~\bibnamefont {Banerjee}},\ }\bibfield
  {title} {\bibinfo {title} {Comparative studies of magnetocaloric effect and
  magnetotransport behavior in {GdRu$_2$Si$_2$} compound},\ }\href
  {https://doi.org/10.1063/1.3043558} {\bibfield  {journal} {\bibinfo
  {journal} {Journal of Applied Physics}\ }\textbf {\bibinfo {volume} {104}},\
  \bibinfo {pages} {123901} (\bibinfo {year} {2008})}\BibitemShut {NoStop}%
\bibitem [{\citenamefont {Garnier}\ \emph {et~al.}(1996)\citenamefont
  {Garnier}, \citenamefont {Gignoux}, \citenamefont {Schmitt},\ and\
  \citenamefont {Shigeoka}}]{Garnier_PB1996}%
  \BibitemOpen
  \bibfield  {author} {\bibinfo {author} {\bibfnamefont {A.}~\bibnamefont
  {Garnier}}, \bibinfo {author} {\bibfnamefont {D.}~\bibnamefont {Gignoux}},
  \bibinfo {author} {\bibfnamefont {D.}~\bibnamefont {Schmitt}},\ and\ \bibinfo
  {author} {\bibfnamefont {T.}~\bibnamefont {Shigeoka}},\ }\bibfield  {title}
  {\bibinfo {title} {Giant magnetic anisotropy in tetragonal {GdRu$_2$Ge$_2$}
  and {GdRu$_2$Si$_2$}},\ }\href
  {https://www.sciencedirect.com/science/article/pii/0921452696000105}
  {\bibfield  {journal} {\bibinfo  {journal} {Physica B: Condensed Matter}\
  }\textbf {\bibinfo {volume} {222}},\ \bibinfo {pages} {80} (\bibinfo {year}
  {1996})}\BibitemShut {NoStop}%
\bibitem [{\citenamefont {Prokle{\v{s}}ka}\ \emph {et~al.}(2006)\citenamefont
  {Prokle{\v{s}}ka}, \citenamefont {Vejpravov{\'a}},\ and\ \citenamefont
  {Sechovsk{\'y}}}]{Prokleska_JP2006}%
  \BibitemOpen
  \bibfield  {author} {\bibinfo {author} {\bibfnamefont {J.}~\bibnamefont
  {Prokle{\v{s}}ka}}, \bibinfo {author} {\bibfnamefont {J.}~\bibnamefont
  {Vejpravov{\'a}}},\ and\ \bibinfo {author} {\bibfnamefont {V.}~\bibnamefont
  {Sechovsk{\'y}}},\ }\bibfield  {title} {\bibinfo {title} {Magnetostriction
  measurement of {GdRu$_2$Si$_2$} single crystal},\ }\href
  {https://doi.org/10.1088/1742-6596/51/1/028} {\bibfield  {journal} {\bibinfo
  {journal} {Journal of Physics: Conference Series}\ }\textbf {\bibinfo
  {volume} {51}},\ \bibinfo {pages} {127} (\bibinfo {year} {2006})}\BibitemShut
  {NoStop}%
\bibitem [{\citenamefont {Bouaziz}\ \emph {et~al.}(2022)\citenamefont
  {Bouaziz}, \citenamefont {Mendive-Tapia}, \citenamefont {Bl\"ugel},\ and\
  \citenamefont {Staunton}}]{Bouaziz_PRL2022}%
  \BibitemOpen
  \bibfield  {author} {\bibinfo {author} {\bibfnamefont {J.}~\bibnamefont
  {Bouaziz}}, \bibinfo {author} {\bibfnamefont {E.}~\bibnamefont
  {Mendive-Tapia}}, \bibinfo {author} {\bibfnamefont {S.}~\bibnamefont
  {Bl\"ugel}},\ and\ \bibinfo {author} {\bibfnamefont {J.~B.}\ \bibnamefont
  {Staunton}},\ }\bibfield  {title} {\bibinfo {title} {Fermi-surface origin of
  skyrmion lattices in centrosymmetric rare-earth intermetallics},\ }\href
  {https://doi.org/10.1103/PhysRevLett.128.157206} {\bibfield  {journal}
  {\bibinfo  {journal} {Phys. Rev. Lett.}\ }\textbf {\bibinfo {volume} {128}},\
  \bibinfo {pages} {157206} (\bibinfo {year} {2022})}\BibitemShut {NoStop}%
\bibitem [{\citenamefont {Yasui}\ \emph {et~al.}(2020)\citenamefont {Yasui},
  \citenamefont {Butler}, \citenamefont {Khanh}, \citenamefont {Hayami},
  \citenamefont {Nomoto}, \citenamefont {Hanaguri}, \citenamefont {Motome},
  \citenamefont {Arita}, \citenamefont {Arima}, \citenamefont {Tokura},\ and\
  \citenamefont {Seki}}]{Yasui_NatComm2020}%
  \BibitemOpen
  \bibfield  {author} {\bibinfo {author} {\bibfnamefont {Y.}~\bibnamefont
  {Yasui}}, \bibinfo {author} {\bibfnamefont {C.~J.}\ \bibnamefont {Butler}},
  \bibinfo {author} {\bibfnamefont {N.~D.}\ \bibnamefont {Khanh}}, \bibinfo
  {author} {\bibfnamefont {S.}~\bibnamefont {Hayami}}, \bibinfo {author}
  {\bibfnamefont {T.}~\bibnamefont {Nomoto}}, \bibinfo {author} {\bibfnamefont
  {T.}~\bibnamefont {Hanaguri}}, \bibinfo {author} {\bibfnamefont
  {Y.}~\bibnamefont {Motome}}, \bibinfo {author} {\bibfnamefont
  {R.}~\bibnamefont {Arita}}, \bibinfo {author} {\bibfnamefont
  {T.}~\bibnamefont {Arima}}, \bibinfo {author} {\bibfnamefont
  {Y.}~\bibnamefont {Tokura}},\ and\ \bibinfo {author} {\bibfnamefont
  {S.}~\bibnamefont {Seki}},\ }\bibfield  {title} {\bibinfo {title} {Imaging
  the coupling between itinerant electrons and localised moments in the
  centrosymmetric skyrmion magnet {GdRu$_2$Si$_2$}},\ }\href
  {https://doi.org/10.1038/s41467-020-19751-4} {\bibfield  {journal} {\bibinfo
  {journal} {Nature Communications}\ }\textbf {\bibinfo {volume} {11}},\
  \bibinfo {pages} {5925} (\bibinfo {year} {2020})}\BibitemShut {NoStop}%
\bibitem [{\citenamefont {Nomoto}\ \emph {et~al.}(2020)\citenamefont {Nomoto},
  \citenamefont {Koretsune},\ and\ \citenamefont {Arita}}]{Nomoto_PRL2020}%
  \BibitemOpen
  \bibfield  {author} {\bibinfo {author} {\bibfnamefont {T.}~\bibnamefont
  {Nomoto}}, \bibinfo {author} {\bibfnamefont {T.}~\bibnamefont {Koretsune}},\
  and\ \bibinfo {author} {\bibfnamefont {R.}~\bibnamefont {Arita}},\ }\bibfield
   {title} {\bibinfo {title} {Formation mechanism of the helical $q$ structure
  in {Gd}-based skyrmion materials},\ }\href
  {https://doi.org/10.1103/PhysRevLett.125.117204} {\bibfield  {journal}
  {\bibinfo  {journal} {Phys. Rev. Lett.}\ }\textbf {\bibinfo {volume} {125}},\
  \bibinfo {pages} {117204} (\bibinfo {year} {2020})}\BibitemShut {NoStop}%
\bibitem [{\citenamefont {Hayami}\ and\ \citenamefont
  {Motome}(2021)}]{Hayami_PRB2021}%
  \BibitemOpen
  \bibfield  {author} {\bibinfo {author} {\bibfnamefont {S.}~\bibnamefont
  {Hayami}}\ and\ \bibinfo {author} {\bibfnamefont {Y.}~\bibnamefont
  {Motome}},\ }\bibfield  {title} {\bibinfo {title} {Square skyrmion crystal in
  centrosymmetric itinerant magnets},\ }\href
  {https://doi.org/10.1103/PhysRevB.103.024439} {\bibfield  {journal} {\bibinfo
   {journal} {Phys. Rev. B}\ }\textbf {\bibinfo {volume} {103}},\ \bibinfo
  {pages} {024439} (\bibinfo {year} {2021})}\BibitemShut {NoStop}%
\bibitem [{\citenamefont {Matsuyama}\ \emph {et~al.}(2023)\citenamefont
  {Matsuyama}, \citenamefont {Nomura}, \citenamefont {Imajo}, \citenamefont
  {Nomoto}, \citenamefont {Arita}, \citenamefont {Sudo}, \citenamefont
  {Kimata}, \citenamefont {Khanh}, \citenamefont {Takagi}, \citenamefont
  {Tokura}, \citenamefont {Seki}, \citenamefont {Kindo},\ and\ \citenamefont
  {Kohama}}]{Matsuyama_PRB23}%
  \BibitemOpen
  \bibfield  {author} {\bibinfo {author} {\bibfnamefont {N.}~\bibnamefont
  {Matsuyama}}, \bibinfo {author} {\bibfnamefont {T.}~\bibnamefont {Nomura}},
  \bibinfo {author} {\bibfnamefont {S.}~\bibnamefont {Imajo}}, \bibinfo
  {author} {\bibfnamefont {T.}~\bibnamefont {Nomoto}}, \bibinfo {author}
  {\bibfnamefont {R.}~\bibnamefont {Arita}}, \bibinfo {author} {\bibfnamefont
  {K.}~\bibnamefont {Sudo}}, \bibinfo {author} {\bibfnamefont {M.}~\bibnamefont
  {Kimata}}, \bibinfo {author} {\bibfnamefont {N.~D.}\ \bibnamefont {Khanh}},
  \bibinfo {author} {\bibfnamefont {R.}~\bibnamefont {Takagi}}, \bibinfo
  {author} {\bibfnamefont {Y.}~\bibnamefont {Tokura}}, \bibinfo {author}
  {\bibfnamefont {S.}~\bibnamefont {Seki}}, \bibinfo {author} {\bibfnamefont
  {K.}~\bibnamefont {Kindo}},\ and\ \bibinfo {author} {\bibfnamefont
  {Y.}~\bibnamefont {Kohama}},\ }\bibfield  {title} {\bibinfo {title} {Quantum
  oscillations in the centrosymmetric skyrmion-hosting magnet
  {GdRu$_2$Si$_2$}},\ }\href {https://doi.org/10.1103/PhysRevB.107.104421}
  {\bibfield  {journal} {\bibinfo  {journal} {Phys. Rev. B}\ }\textbf {\bibinfo
  {volume} {107}},\ \bibinfo {pages} {104421} (\bibinfo {year}
  {2023})}\BibitemShut {NoStop}%
\bibitem [{\citenamefont {Bihlmayer}\ \emph {et~al.}(2022)\citenamefont
  {Bihlmayer}, \citenamefont {Noel}, \citenamefont {Vyalikh}, \citenamefont
  {Chulkov},\ and\ \citenamefont {Manchon}}]{Gustav2022}%
  \BibitemOpen
  \bibfield  {author} {\bibinfo {author} {\bibfnamefont {G.}~\bibnamefont
  {Bihlmayer}}, \bibinfo {author} {\bibfnamefont {P.}~\bibnamefont {Noel}},
  \bibinfo {author} {\bibfnamefont {D.~V.}\ \bibnamefont {Vyalikh}}, \bibinfo
  {author} {\bibfnamefont {E.~V.}\ \bibnamefont {Chulkov}},\ and\ \bibinfo
  {author} {\bibfnamefont {A.}~\bibnamefont {Manchon}},\ }\bibfield  {title}
  {\bibinfo {title} {Rashba-like physics in condensed matter},\ }\href
  {https://doi.org/10.1038/s42254-022-00490-y} {\bibfield  {journal} {\bibinfo
  {journal} {Nature Reviews Physics}\ }\textbf {\bibinfo {volume} {4}},\
  \bibinfo {pages} {642} (\bibinfo {year} {2022})}\BibitemShut {NoStop}%
\bibitem [{\citenamefont {Liechtenstein}\ \emph {et~al.}(1987)\citenamefont
  {Liechtenstein}, \citenamefont {Katsnelson}, \citenamefont {Antropov},\ and\
  \citenamefont {Gubanov}}]{Liechtenstein1987}%
  \BibitemOpen
  \bibfield  {author} {\bibinfo {author} {\bibfnamefont {A.~I.}\ \bibnamefont
  {Liechtenstein}}, \bibinfo {author} {\bibfnamefont {M.~I.}\ \bibnamefont
  {Katsnelson}}, \bibinfo {author} {\bibfnamefont {V.~P.}\ \bibnamefont
  {Antropov}},\ and\ \bibinfo {author} {\bibfnamefont {V.~A.}\ \bibnamefont
  {Gubanov}},\ }\bibfield  {title} {\bibinfo {title} {Local spin density
  functional approach to the theory of exchange interactions in ferromagnetic
  metals and alloys},\ }\href {https://doi.org/10.1016/0304-8853(87)90721-9}
  {\bibfield  {journal} {\bibinfo  {journal} {Journal of Magnetism and Magnetic
  Materials}\ }\textbf {\bibinfo {volume} {67}},\ \bibinfo {pages} {65 }
  (\bibinfo {year} {1987})}\BibitemShut {NoStop}%
\bibitem [{\citenamefont {Hoffmann}\ \emph {et~al.}(2020)\citenamefont
  {Hoffmann}, \citenamefont {Ernst}, \citenamefont {Hergert}, \citenamefont
  {Antonov}, \citenamefont {Adeagbo}, \citenamefont {Geilhufe},\ and\
  \citenamefont {Ben~Hamed}}]{Hoffmann2020}%
  \BibitemOpen
  \bibfield  {author} {\bibinfo {author} {\bibfnamefont {M.}~\bibnamefont
  {Hoffmann}}, \bibinfo {author} {\bibfnamefont {A.}~\bibnamefont {Ernst}},
  \bibinfo {author} {\bibfnamefont {W.}~\bibnamefont {Hergert}}, \bibinfo
  {author} {\bibfnamefont {V.~N.}\ \bibnamefont {Antonov}}, \bibinfo {author}
  {\bibfnamefont {W.~A.}\ \bibnamefont {Adeagbo}}, \bibinfo {author}
  {\bibfnamefont {R.~M.}\ \bibnamefont {Geilhufe}},\ and\ \bibinfo {author}
  {\bibfnamefont {H.}~\bibnamefont {Ben~Hamed}},\ }\bibfield  {title} {\bibinfo
  {title} {Magnetic and electronic properties of complex oxides from
  first-principles},\ }\href
  {https://doi.org/https://doi.org/10.1002/pssb.201900671} {\bibfield
  {journal} {\bibinfo  {journal} {Physica Status Solidi (b)}\ }\textbf
  {\bibinfo {volume} {257}},\ \bibinfo {pages} {1900671} (\bibinfo {year}
  {2020})}\BibitemShut {NoStop}%
\bibitem [{\citenamefont {G{\"u}ttler}\ \emph {et~al.}(2016)\citenamefont
  {G{\"u}ttler}, \citenamefont {Generalov}, \citenamefont {Otrokov},
  \citenamefont {Kummer}, \citenamefont {Kliemt}, \citenamefont {Fedorov},
  \citenamefont {Chikina}, \citenamefont {Danzenb{\"a}cher}, \citenamefont
  {Schulz}, \citenamefont {Chulkov}, \citenamefont {Koroteev}, \citenamefont
  {Caroca-Canales}, \citenamefont {Shi}, \citenamefont {Radovic}, \citenamefont
  {Geibel}, \citenamefont {Laubschat}, \citenamefont {Dudin}, \citenamefont
  {Kim}, \citenamefont {Hoesch}, \citenamefont {Krellner},\ and\ \citenamefont
  {Vyalikh}}]{Guttler2016}%
  \BibitemOpen
  \bibfield  {author} {\bibinfo {author} {\bibfnamefont {M.}~\bibnamefont
  {G{\"u}ttler}}, \bibinfo {author} {\bibfnamefont {A.}~\bibnamefont
  {Generalov}}, \bibinfo {author} {\bibfnamefont {M.~M.}\ \bibnamefont
  {Otrokov}}, \bibinfo {author} {\bibfnamefont {K.}~\bibnamefont {Kummer}},
  \bibinfo {author} {\bibfnamefont {K.}~\bibnamefont {Kliemt}}, \bibinfo
  {author} {\bibfnamefont {A.}~\bibnamefont {Fedorov}}, \bibinfo {author}
  {\bibfnamefont {A.}~\bibnamefont {Chikina}}, \bibinfo {author} {\bibfnamefont
  {S.}~\bibnamefont {Danzenb{\"a}cher}}, \bibinfo {author} {\bibfnamefont
  {S.}~\bibnamefont {Schulz}}, \bibinfo {author} {\bibfnamefont {E.~V.}\
  \bibnamefont {Chulkov}}, \bibinfo {author} {\bibfnamefont {Y.~M.}\
  \bibnamefont {Koroteev}}, \bibinfo {author} {\bibfnamefont {N.}~\bibnamefont
  {Caroca-Canales}}, \bibinfo {author} {\bibfnamefont {M.}~\bibnamefont {Shi}},
  \bibinfo {author} {\bibfnamefont {M.}~\bibnamefont {Radovic}}, \bibinfo
  {author} {\bibfnamefont {C.}~\bibnamefont {Geibel}}, \bibinfo {author}
  {\bibfnamefont {C.}~\bibnamefont {Laubschat}}, \bibinfo {author}
  {\bibfnamefont {P.}~\bibnamefont {Dudin}}, \bibinfo {author} {\bibfnamefont
  {T.~K.}\ \bibnamefont {Kim}}, \bibinfo {author} {\bibfnamefont
  {M.}~\bibnamefont {Hoesch}}, \bibinfo {author} {\bibfnamefont
  {C.}~\bibnamefont {Krellner}},\ and\ \bibinfo {author} {\bibfnamefont
  {D.~V.}\ \bibnamefont {Vyalikh}},\ }\bibfield  {title} {\bibinfo {title}
  {Robust and tunable itinerant ferromagnetism at the silicon surface of the
  antiferromagnet {GdRh$_2$Si$_2$}},\ }\href
  {https://doi.org/10.1038/srep24254} {\bibfield  {journal} {\bibinfo
  {journal} {Scientific Reports}\ }\textbf {\bibinfo {volume} {6}},\ \bibinfo
  {pages} {24254} (\bibinfo {year} {2016})}\BibitemShut {NoStop}%
\bibitem [{\citenamefont {Kliemt}\ \emph {et~al.}(2020)\citenamefont {Kliemt},
  \citenamefont {Peters}, \citenamefont {Feldmann}, \citenamefont {Kraiker},
  \citenamefont {Tran}, \citenamefont {Rongstock}, \citenamefont {Hellwig},
  \citenamefont {Witt}, \citenamefont {Bolte},\ and\ \citenamefont
  {Krellner}}]{Kliemt2019}%
  \BibitemOpen
  \bibfield  {author} {\bibinfo {author} {\bibfnamefont {K.}~\bibnamefont
  {Kliemt}}, \bibinfo {author} {\bibfnamefont {M.}~\bibnamefont {Peters}},
  \bibinfo {author} {\bibfnamefont {F.}~\bibnamefont {Feldmann}}, \bibinfo
  {author} {\bibfnamefont {A.}~\bibnamefont {Kraiker}}, \bibinfo {author}
  {\bibfnamefont {D.-M.}\ \bibnamefont {Tran}}, \bibinfo {author}
  {\bibfnamefont {S.}~\bibnamefont {Rongstock}}, \bibinfo {author}
  {\bibfnamefont {J.}~\bibnamefont {Hellwig}}, \bibinfo {author} {\bibfnamefont
  {S.}~\bibnamefont {Witt}}, \bibinfo {author} {\bibfnamefont {M.}~\bibnamefont
  {Bolte}},\ and\ \bibinfo {author} {\bibfnamefont {C.}~\bibnamefont
  {Krellner}},\ }\bibfield  {title} {\bibinfo {title} {Crystal growth of
  materials with the {ThCr$_2$Si$_2$} structure type},\ }\href
  {https://doi.org/https://doi.org/10.1002/crat.201900116} {\bibfield
  {journal} {\bibinfo  {journal} {Cryst. Res. Technol.}\ }\textbf {\bibinfo
  {volume} {55}},\ \bibinfo {pages} {1900116} (\bibinfo {year}
  {2020})}\BibitemShut {NoStop}%
\bibitem [{\citenamefont {Bl\"ochl}(1994{\natexlab{a}})}]{Blochl.prb1994}%
  \BibitemOpen
  \bibfield  {author} {\bibinfo {author} {\bibfnamefont {P.~E.}\ \bibnamefont
  {Bl\"ochl}},\ }\bibfield  {title} {\bibinfo {title} {Projector augmented-wave
  method},\ }\href@noop {} {\bibfield  {journal} {\bibinfo  {journal} {Phys.
  Rev. B}\ }\textbf {\bibinfo {volume} {50}},\ \bibinfo {pages} {17953}
  (\bibinfo {year} {1994}{\natexlab{a}})}\BibitemShut {NoStop}%
\bibitem [{\citenamefont {Bl\"ochl}(1994{\natexlab{b}})}]{PAW1}%
  \BibitemOpen
  \bibfield  {author} {\bibinfo {author} {\bibfnamefont {P.~E.}\ \bibnamefont
  {Bl\"ochl}},\ }\bibfield  {title} {\bibinfo {title} {Projector augmented-wave
  method},\ }\href {https://doi.org/10.1103/PhysRevB.50.17953} {\bibfield
  {journal} {\bibinfo  {journal} {Phys. Rev. B}\ }\textbf {\bibinfo {volume}
  {50}},\ \bibinfo {pages} {17953} (\bibinfo {year}
  {1994}{\natexlab{b}})}\BibitemShut {NoStop}%
\bibitem [{\citenamefont {Kresse}\ and\ \citenamefont
  {Joubert}(1999{\natexlab{a}})}]{PAW2}%
  \BibitemOpen
  \bibfield  {author} {\bibinfo {author} {\bibfnamefont {G.}~\bibnamefont
  {Kresse}}\ and\ \bibinfo {author} {\bibfnamefont {D.}~\bibnamefont
  {Joubert}},\ }\bibfield  {title} {\bibinfo {title} {From ultrasoft
  pseudopotentials to the projector augmented-wave method},\ }\href
  {https://doi.org/10.1103/PhysRevB.59.1758} {\bibfield  {journal} {\bibinfo
  {journal} {Phys. Rev. B}\ }\textbf {\bibinfo {volume} {59}},\ \bibinfo
  {pages} {1758} (\bibinfo {year} {1999}{\natexlab{a}})}\BibitemShut {NoStop}%
\bibitem [{\citenamefont {Kresse}\ and\ \citenamefont
  {Furthm\"uller}(1996)}]{vasp1}%
  \BibitemOpen
  \bibfield  {author} {\bibinfo {author} {\bibfnamefont {G.}~\bibnamefont
  {Kresse}}\ and\ \bibinfo {author} {\bibfnamefont {J.}~\bibnamefont
  {Furthm\"uller}},\ }\bibfield  {title} {\bibinfo {title} {Efficient iterative
  schemes for ab initio total-energy calculations using a plane-wave basis
  set},\ }\href@noop {} {\bibfield  {journal} {\bibinfo  {journal} {Phys. Rev.
  B}\ }\textbf {\bibinfo {volume} {54}},\ \bibinfo {pages} {11169} (\bibinfo
  {year} {1996})}\BibitemShut {NoStop}%
\bibitem [{\citenamefont {Kresse}\ and\ \citenamefont
  {Joubert}(1999{\natexlab{b}})}]{vasp2}%
  \BibitemOpen
  \bibfield  {author} {\bibinfo {author} {\bibfnamefont {G.}~\bibnamefont
  {Kresse}}\ and\ \bibinfo {author} {\bibfnamefont {D.}~\bibnamefont
  {Joubert}},\ }\bibfield  {title} {\bibinfo {title} {From ultrasoft
  pseudopotentials to the projector augmented-wave method},\ }\href@noop {}
  {\bibfield  {journal} {\bibinfo  {journal} {Phys. Rev. B}\ }\textbf {\bibinfo
  {volume} {59}},\ \bibinfo {pages} {1758} (\bibinfo {year}
  {1999}{\natexlab{b}})}\BibitemShut {NoStop}%
\bibitem [{\citenamefont {Perdew}\ \emph {et~al.}(1996)\citenamefont {Perdew},
  \citenamefont {Burke},\ and\ \citenamefont {Ernzerhof}}]{Perdew.prl1996}%
  \BibitemOpen
  \bibfield  {author} {\bibinfo {author} {\bibfnamefont {J.~P.}\ \bibnamefont
  {Perdew}}, \bibinfo {author} {\bibfnamefont {K.}~\bibnamefont {Burke}},\ and\
  \bibinfo {author} {\bibfnamefont {M.}~\bibnamefont {Ernzerhof}},\ }\bibfield
  {title} {\bibinfo {title} {Generalized gradient approximation made simple},\
  }\href@noop {} {\bibfield  {journal} {\bibinfo  {journal} {Phys. Rev. Lett.}\
  }\textbf {\bibinfo {volume} {77}},\ \bibinfo {pages} {3865} (\bibinfo {year}
  {1996})}\BibitemShut {NoStop}%
\bibitem [{\citenamefont {Krukau}\ \emph {et~al.}(2006)\citenamefont {Krukau},
  \citenamefont {Vydrov}, \citenamefont {Izmaylov},\ and\ \citenamefont
  {Scuseria}}]{HSE06}%
  \BibitemOpen
  \bibfield  {author} {\bibinfo {author} {\bibfnamefont {A.~V.}\ \bibnamefont
  {Krukau}}, \bibinfo {author} {\bibfnamefont {O.~A.}\ \bibnamefont {Vydrov}},
  \bibinfo {author} {\bibfnamefont {A.~F.}\ \bibnamefont {Izmaylov}},\ and\
  \bibinfo {author} {\bibfnamefont {G.~E.}\ \bibnamefont {Scuseria}},\
  }\bibfield  {title} {\bibinfo {title} {Influence of the exchange screening
  parameter on the performance of screened hybrid functionals},\ }\href
  {https://doi.org/10.1063/1.2404663} {\bibfield  {journal} {\bibinfo
  {journal} {The Journal of Chemical Physics}\ }\textbf {\bibinfo {volume}
  {125}},\ \bibinfo {pages} {224106} (\bibinfo {year} {2006})}\BibitemShut
  {NoStop}%
\bibitem [{\citenamefont {Anisimov}\ \emph {et~al.}(1991)\citenamefont
  {Anisimov}, \citenamefont {Zaanen},\ and\ \citenamefont
  {Andersen}}]{Anisimov1991}%
  \BibitemOpen
  \bibfield  {author} {\bibinfo {author} {\bibfnamefont {V.~I.}\ \bibnamefont
  {Anisimov}}, \bibinfo {author} {\bibfnamefont {J.}~\bibnamefont {Zaanen}},\
  and\ \bibinfo {author} {\bibfnamefont {O.~K.}\ \bibnamefont {Andersen}},\
  }\bibfield  {title} {\bibinfo {title} {Band theory and mott insulators:
  {Hubbard $U$} instead of {Stoner $I$}},\ }\href
  {https://doi.org/10.1103/PhysRevB.44.943} {\bibfield  {journal} {\bibinfo
  {journal} {Phys. Rev. B}\ }\textbf {\bibinfo {volume} {44}},\ \bibinfo
  {pages} {943} (\bibinfo {year} {1991})}\BibitemShut {NoStop}%
\bibitem [{\citenamefont {Ferreira}\ \emph {et~al.}(2008)\citenamefont
  {Ferreira}, \citenamefont {Marques},\ and\ \citenamefont {Teles}}]{DFT12_1}%
  \BibitemOpen
  \bibfield  {author} {\bibinfo {author} {\bibfnamefont {L.~G.}\ \bibnamefont
  {Ferreira}}, \bibinfo {author} {\bibfnamefont {M.}~\bibnamefont {Marques}},\
  and\ \bibinfo {author} {\bibfnamefont {L.~K.}\ \bibnamefont {Teles}},\
  }\bibfield  {title} {\bibinfo {title} {Approximation to density functional
  theory for the calculation of band gaps of semiconductors},\ }\href
  {https://doi.org/10.1103/PhysRevB.78.125116} {\bibfield  {journal} {\bibinfo
  {journal} {Phys. Rev. B}\ }\textbf {\bibinfo {volume} {78}},\ \bibinfo
  {pages} {125116} (\bibinfo {year} {2008})}\BibitemShut {NoStop}%
\bibitem [{\citenamefont {Ferreira}\ \emph {et~al.}(2011)\citenamefont
  {Ferreira}, \citenamefont {Marques},\ and\ \citenamefont {Teles}}]{DFT12_2}%
  \BibitemOpen
  \bibfield  {author} {\bibinfo {author} {\bibfnamefont {L.~G.}\ \bibnamefont
  {Ferreira}}, \bibinfo {author} {\bibfnamefont {M.}~\bibnamefont {Marques}},\
  and\ \bibinfo {author} {\bibfnamefont {L.~K.}\ \bibnamefont {Teles}},\
  }\bibfield  {title} {\bibinfo {title} {Slater half-occupation technique
  revisited: the {LDA-1/2} and {GGA-1/2} approaches for atomic ionization
  energies and band gaps in semiconductors},\ }\href
  {https://doi.org/10.1063/1.3624562} {\bibfield  {journal} {\bibinfo
  {journal} {AIP Advances}\ }\textbf {\bibinfo {volume} {1}},\ \bibinfo {pages}
  {032119} (\bibinfo {year} {2011})}\BibitemShut {NoStop}%
\bibitem [{\citenamefont {Sandratskii}(1998)}]{Sandratskii1998}%
  \BibitemOpen
  \bibfield  {author} {\bibinfo {author} {\bibfnamefont {L.~M.}\ \bibnamefont
  {Sandratskii}},\ }\bibfield  {title} {\bibinfo {title} {Noncollinear
  magnetism in itinerant-electron systems: Theory and applications},\ }\href
  {https://doi.org/10.1080/000187398243573} {\bibfield  {journal} {\bibinfo
  {journal} {Advances in Physics}\ }\textbf {\bibinfo {volume} {47}},\ \bibinfo
  {pages} {91} (\bibinfo {year} {1998})}\BibitemShut {NoStop}%
\bibitem [{\citenamefont {Medeiros}\ \emph {et~al.}(2014)\citenamefont
  {Medeiros}, \citenamefont {Stafstr\"om},\ and\ \citenamefont
  {Bj\"ork}}]{Unfolding-1-PRB-2014}%
  \BibitemOpen
  \bibfield  {author} {\bibinfo {author} {\bibfnamefont {P.~V.~C.}\
  \bibnamefont {Medeiros}}, \bibinfo {author} {\bibfnamefont {S.}~\bibnamefont
  {Stafstr\"om}},\ and\ \bibinfo {author} {\bibfnamefont {J.}~\bibnamefont
  {Bj\"ork}},\ }\bibfield  {title} {\bibinfo {title} {Effects of extrinsic and
  intrinsic perturbations on the electronic structure of graphene: Retaining an
  effective primitive cell band structure by band unfolding},\ }\href
  {https://doi.org/10.1103/PhysRevB.89.041407} {\bibfield  {journal} {\bibinfo
  {journal} {Phys. Rev. B}\ }\textbf {\bibinfo {volume} {89}},\ \bibinfo
  {pages} {041407} (\bibinfo {year} {2014})}\BibitemShut {NoStop}%
\bibitem [{\citenamefont {Medeiros}\ \emph {et~al.}(2015)\citenamefont
  {Medeiros}, \citenamefont {Tsirkin}, \citenamefont {Stafstr\"om},\ and\
  \citenamefont {Bj\"ork}}]{Unfolding-2-PRB-2015}%
  \BibitemOpen
  \bibfield  {author} {\bibinfo {author} {\bibfnamefont {P.~V.~C.}\
  \bibnamefont {Medeiros}}, \bibinfo {author} {\bibfnamefont {S.~S.}\
  \bibnamefont {Tsirkin}}, \bibinfo {author} {\bibfnamefont {S.}~\bibnamefont
  {Stafstr\"om}},\ and\ \bibinfo {author} {\bibfnamefont {J.}~\bibnamefont
  {Bj\"ork}},\ }\bibfield  {title} {\bibinfo {title} {Unfolding spinor wave
  functions and expectation values of general operators: Introducing the
  unfolding-density operator},\ }\href
  {https://doi.org/10.1103/PhysRevB.91.041116} {\bibfield  {journal} {\bibinfo
  {journal} {Phys. Rev. B}\ }\textbf {\bibinfo {volume} {91}},\ \bibinfo
  {pages} {041116} (\bibinfo {year} {2015})}\BibitemShut {NoStop}%
\bibitem [{\citenamefont {Koepernik}\ and\ \citenamefont
  {Eschrig}(1999)}]{FPLO}%
  \BibitemOpen
  \bibfield  {author} {\bibinfo {author} {\bibfnamefont {K.}~\bibnamefont
  {Koepernik}}\ and\ \bibinfo {author} {\bibfnamefont {H.}~\bibnamefont
  {Eschrig}},\ }\bibfield  {title} {\bibinfo {title} {{Full-Potential
  Nonorthogonal Local-Orbital Minimum-Basis Band-Structure Scheme}},\ }\href
  {https://doi.org/10.1103/PhysRevB.59.1743} {\bibfield  {journal} {\bibinfo
  {journal} {Phys. Rev. B}\ }\textbf {\bibinfo {volume} {59}},\ \bibinfo
  {pages} {1743} (\bibinfo {year} {1999})}\BibitemShut {NoStop}%
\bibitem [{\citenamefont {Momma}\ and\ \citenamefont {Izumi}(2011)}]{VESTA}%
  \BibitemOpen
  \bibfield  {author} {\bibinfo {author} {\bibfnamefont {K.}~\bibnamefont
  {Momma}}\ and\ \bibinfo {author} {\bibfnamefont {F.}~\bibnamefont {Izumi}},\
  }\bibfield  {title} {\bibinfo {title} {{\it VESTA 3} for three-dimensional
  visualization of crystal, volumetric and morphology data},\ }\href
  {https://doi.org/10.1107/S0021889811038970} {\bibfield  {journal} {\bibinfo
  {journal} {Journal of Applied Crystallography}\ }\textbf {\bibinfo {volume}
  {44}},\ \bibinfo {pages} {1272} (\bibinfo {year} {2011})}\BibitemShut
  {NoStop}%
\bibitem [{\citenamefont {Kawamura}(2019)}]{Kawamura2019}%
  \BibitemOpen
  \bibfield  {author} {\bibinfo {author} {\bibfnamefont {M.}~\bibnamefont
  {Kawamura}},\ }\bibfield  {title} {\bibinfo {title} {Fermisurfer:
  Fermi-surface viewer providing multiple representation schemes},\ }\href
  {https://www.sciencedirect.com/science/article/pii/S0010465519300347}
  {\bibfield  {journal} {\bibinfo  {journal} {Computer Physics Communications}\
  }\textbf {\bibinfo {volume} {239}},\ \bibinfo {pages} {197} (\bibinfo {year}
  {2019})}\BibitemShut {NoStop}%
\bibitem [{\citenamefont {Dudarev}\ \emph {et~al.}(1998)\citenamefont
  {Dudarev}, \citenamefont {Botton}, \citenamefont {Savrasov}, \citenamefont
  {Humphreys},\ and\ \citenamefont {Sutton}}]{Dudarev.prb1998}%
  \BibitemOpen
  \bibfield  {author} {\bibinfo {author} {\bibfnamefont {S.~L.}\ \bibnamefont
  {Dudarev}}, \bibinfo {author} {\bibfnamefont {G.~A.}\ \bibnamefont {Botton}},
  \bibinfo {author} {\bibfnamefont {S.~Y.}\ \bibnamefont {Savrasov}}, \bibinfo
  {author} {\bibfnamefont {C.~J.}\ \bibnamefont {Humphreys}},\ and\ \bibinfo
  {author} {\bibfnamefont {A.~P.}\ \bibnamefont {Sutton}},\ }\bibfield  {title}
  {\bibinfo {title} {Electron-energy-loss spectra and the structural stability
  of nickel oxide: An {LSDA+$U$} study},\ }\href@noop {} {\bibfield  {journal}
  {\bibinfo  {journal} {Phys. Rev. B}\ }\textbf {\bibinfo {volume} {57}},\
  \bibinfo {pages} {1505} (\bibinfo {year} {1998})}\BibitemShut {NoStop}%
\bibitem [{\citenamefont {Perdew}\ and\ \citenamefont
  {Zunger}(1981)}]{Perdew1981}%
  \BibitemOpen
  \bibfield  {author} {\bibinfo {author} {\bibfnamefont {J.~P.}\ \bibnamefont
  {Perdew}}\ and\ \bibinfo {author} {\bibfnamefont {A.}~\bibnamefont
  {Zunger}},\ }\bibfield  {title} {\bibinfo {title} {Self-interaction
  correction to density-functional approximations for many-electron systems},\
  }\href {https://doi.org/10.1103/PhysRevB.23.5048} {\bibfield  {journal}
  {\bibinfo  {journal} {Phys. Rev. B}\ }\textbf {\bibinfo {volume} {23}},\
  \bibinfo {pages} {5048} (\bibinfo {year} {1981})}\BibitemShut {NoStop}%
\bibitem [{\citenamefont {L\"uders}\ \emph {et~al.}(2005)\citenamefont
  {L\"uders}, \citenamefont {Ernst}, \citenamefont {D\"ane}, \citenamefont
  {Szotek}, \citenamefont {Svane}, \citenamefont {K\"odderitzsch},
  \citenamefont {Hergert}, \citenamefont {Gy\"orffy},\ and\ \citenamefont
  {Temmerman}}]{Luders2005}%
  \BibitemOpen
  \bibfield  {author} {\bibinfo {author} {\bibfnamefont {M.}~\bibnamefont
  {L\"uders}}, \bibinfo {author} {\bibfnamefont {A.}~\bibnamefont {Ernst}},
  \bibinfo {author} {\bibfnamefont {M.}~\bibnamefont {D\"ane}}, \bibinfo
  {author} {\bibfnamefont {Z.}~\bibnamefont {Szotek}}, \bibinfo {author}
  {\bibfnamefont {A.}~\bibnamefont {Svane}}, \bibinfo {author} {\bibfnamefont
  {D.}~\bibnamefont {K\"odderitzsch}}, \bibinfo {author} {\bibfnamefont
  {W.}~\bibnamefont {Hergert}}, \bibinfo {author} {\bibfnamefont {B.~L.}\
  \bibnamefont {Gy\"orffy}},\ and\ \bibinfo {author} {\bibfnamefont {W.~M.}\
  \bibnamefont {Temmerman}},\ }\bibfield  {title} {\bibinfo {title}
  {Self-interaction correction in multiple scattering theory},\ }\href
  {https://doi.org/10.1103/PhysRevB.71.205109} {\bibfield  {journal} {\bibinfo
  {journal} {Phys. Rev. B}\ }\textbf {\bibinfo {volume} {71}},\ \bibinfo
  {pages} {205109} (\bibinfo {year} {2005})}\BibitemShut {NoStop}%
\end{thebibliography}

%

\end{document}